\documentclass[%
 aip,
 pof,%
 amsmath,amssymb,
 reprint,%
]{revtex4-1}
\usepackage {CJK}
\usepackage{graphicx}
\usepackage{dcolumn}
\usepackage{bm}
\usepackage{color}
\usepackage{hyperref}

\newcommand\We{\mbox{\textit{We}}}  
\usepackage{makecell}
\begin{document}
\begin{CJK*}{GB}{gbsn}
\preprint{AIP/123-QED}

\title{Numerical simulation of secondary breakup of shear-thinning droplets}
\affiliation{State Key Laboratory of Engines, Tianjin University, Tianjin, 300072, China.}%
\author{Yang Li (李杨)}
\affiliation{State Key Laboratory of Engines, Tianjin University, Tianjin, 300072, China.}%
\author{Zhikun Xu (徐志坤)}
\author{Xiaoyun Peng (彭小芸)}
\affiliation{State Key Laboratory of Engines, Tianjin University, Tianjin, 300072, China.}%
\author{Tianyou Wang (王天友)}
\affiliation{State Key Laboratory of Engines, Tianjin University, Tianjin, 300072, China.}%
\author{Zhizhao Che (车志钊)}
\email{chezhizhao@tju.edu.cn}
 \affiliation{State Key Laboratory of Engines, Tianjin University, Tianjin, 300072, China.}

\date{\today}

\begin{abstract}
The breakup of non-Newtonian droplets is ubiquitous in numerous applications. Although the non-Newtonian property can significantly change the droplet breakup process, most previous studies consider Newtonian droplets, and the effects of the non-Newtonian properties on the breakup process are still unclear. This study focuses on the secondary breakup of shear-thinning droplets by numerical simulation. The volume of fluid method is used to capture interface dynamics on adaptive grids. To compare shear-thinning droplets and Newtonian droplets, a new definition of the Ohnesorge number is proposed by considering the characteristic shear rate in the droplet induced by the airflow. The results show that compared with the Newtonian fluid, the shear-thinning properties can change the effective viscosity distribution inside the droplet, alter the local deformation, change the droplet morphology, and affect the transition in the droplet breakup regime.
\end{abstract}


\maketitle
\end{CJK*}

\section{Introduction}\label{sec:sec01}
The breakup of droplets is a ubiquitous process in nature and many industrial applications, such as raindrop formation, fuel injection, spray cooling, and powder production. Compared with Newtonian fluids, the characteristics of non-Newtonian fluids have significant effects on the deformation and breakup of droplets. As a common type of non-Newtonian fluids, shear-thinning fluids are widely used in many applications, for example, the atomization of gel fuel in engines. Gel fuel is a type of fuel that can be obtained by gelling traditional fuels. It generally exhibits a shear-thinning property \cite{Natan2002GelPropellants, Manisha2021Gel} and has great potential for rocket propulsion. However, the shear-thinning property of gel fuel makes the atomization process complex. To utilize gel fuel effectively, more efforts are needed to understand its atomization process. During the atomization, secondary breakup is a process that droplets further deform and break up into smaller droplets under aerodynamic force. The droplet breakup directly affects the final droplet size of a spray, which further affects the ignition delay and the combustion efficiency in fuel combustion processes. In addition, the shear-thinning property is common in polymer solutions, complex fluids, and suspensions. The breakup of shear-thinning droplets is also involved in many other applications, such as spray coating of thin-film catalyst layers and spray drying in material synthesis. Therefore, it is important to study the breakup of shear-thinning droplets for the relevant applications.

To understand droplet dynamics in the secondary breakup, many researchers have carried out experimental studies using Newtonian fluids \cite{Cao2007Identified, Dai2001PulsedHolography, Joseph2002RayleighTaylor, Kulkarni2014BagBreakup, Opfer2014FilmThickness, Xu2020ShearFlow, Zhao2010MorphologicalClassification, Zhao2013TemporalProperties, Zhang2016Surfactant}. Numerical simulations also play an important role in the study of secondary breakup \cite{Dorschner2020TransverseRTinstability, Feng2010InterfacialFlows, Jain2015HighDensityRatio, Jain2019ModerateWeber, Jiao2019TurbulentFlows, Ling2021LateralPulsating, Marcotte2019Thresholds, Meng2018SheetInstability, Sojka2021MultimodeBreakup, Rimbert2020DropletDeformation, Stefanitsis2019DropletsInTandem, Yang2017Transitions, Yang2016HighlyUnstable, Zhu2021AirflowPressure}. It can provide detailed information about the breakup process, which is difficult to measure directly in experiments. It is also helpful to understand droplet breakup in different conditions such as high-density ratio \cite{Jain2019ModerateWeber}, pulsating airflow \cite{Ling2021LateralPulsating}, and airflow pressure \cite{Zhu2021AirflowPressure}. For example, numerical simulations have shown that the liquid-gas density ratio significantly affects the transition state of the critical Weber number and the breakup mode \cite{Marcotte2019Thresholds, Yang2017Transitions, Yang2016HighlyUnstable}. The change in the Reynolds number causes significant changes in droplet acceleration, deformation, and breakup dynamics \cite{Jain2019ModerateWeber}.

For non-Newtonian fluids, the mechanism of secondary droplet breakup has been studied via experiments \cite{Gao2014Inline, Qian2021CarboxymethylCelluloseDroplets, Snyder2010ElasticNonNewtonian, Wang2021ShearThickening, Zhao2011CoalWaterSlurry} and numerical simulations \cite{Cao2022KeroseneGel, Chu2020PolymerSolution, Fu2019ImpactModel, Kant2022BreakupMechanism, Markovich2019Destruction, Minakov2019Petrochemicals, Tavangar2015CoalWaterSlurry, Verhulst2009BlendMorphology, Wong2019DispersedPhase}. These studies typically focused on the power-law rheological model \cite{Cao2022KeroseneGel, Kant2022BreakupMechanism, Qian2021CarboxymethylCelluloseDroplets, Wang2021ShearThickening}, Bingham rheological model \cite{Markovich2019Destruction, Minakov2019Petrochemicals, Tavangar2015CoalWaterSlurry, Zhao2011CoalWaterSlurry}, Herschel-Bulkley rheological model \cite{Chu2020PolymerSolution}, and Oldroyd-B rheological model \cite{Snyder2010ElasticNonNewtonian, Verhulst2009BlendMorphology}. The results showed that the non-Newtonian characteristics are an important factor affecting the droplet breakup process. For the shear-thickening property, it can induce a hardening deformation mode when the droplet enters the airflow field \cite{Wang2021ShearThickening}. For the shear-thinning property, it can make the liquid ring and liquid segments generated by the droplet unable to be completely fragmented under aerodynamic force \cite{Qian2021CarboxymethylCelluloseDroplets}, and there is a significant change in the deformation rate near the critical Weber number \cite{Cao2022KeroseneGel}. For polymer solution, Herschel-Bulkley constitutive equations can be used to describe the shear-thinning behavior, and the liquid bag will form an obvious reticular structure which is very different from the breakup of Newtonian fluids \cite{Chu2020PolymerSolution}. Coal water slurry can be treated as Bingham liquid approximately. Based on the morphology, deformation, and breakup regimes, the breakup of coal water slurry droplets can be categorized into hole breakup and tensile breakup modes \cite{Markovich2019Destruction, Minakov2019Petrochemicals}. For viscoelastic droplets, apparent persistent ligaments are formed during the bag breakup, rather than being broken directly into smaller droplets for Newtonian droplets \cite{Snyder2010ElasticNonNewtonian}. Although it is widely accepted that the non-Newtonian property can significantly change the droplet breakup process, the above studies mainly focus on the morphology and the difficulty of droplet deformation and breakup, but the mechanism of difference is still not clear, for example, the reason of the unique deformation characteristics of the non-Newtonian droplets.

In this study of the breakup of droplets, we focus on the influence of the shear-thinning property on the breakup process. We find that the shear-thinning property can significantly affect the droplet breakup process by inducing unique viscosity distribution within the droplet. The volume of fluid (VOF) method and adaptive mesh refinement (AMR) technique are used to capture the gas-liquid interface. To compare shear-thinning droplets and Newtonian droplets, a new definition of the Ohnesorge number is proposed by considering the characteristic shear rate in the droplet induced by the airflow. The numerical results are compared with the experimental results to verify the numerical model. The results of shear-thinning droplets are compared with that of Newtonian droplets. Finally, the effects of rheological parameters on the droplet deformation and the breakup mode are analyzed by considering the distribution of the effective viscosity in the droplet.

\section{Models and methods}\label{sec:sec02}
\subsection{Problem description}\label{sec:sec021}
\begin{figure}
  \centerline{\includegraphics[width=0.7\columnwidth]{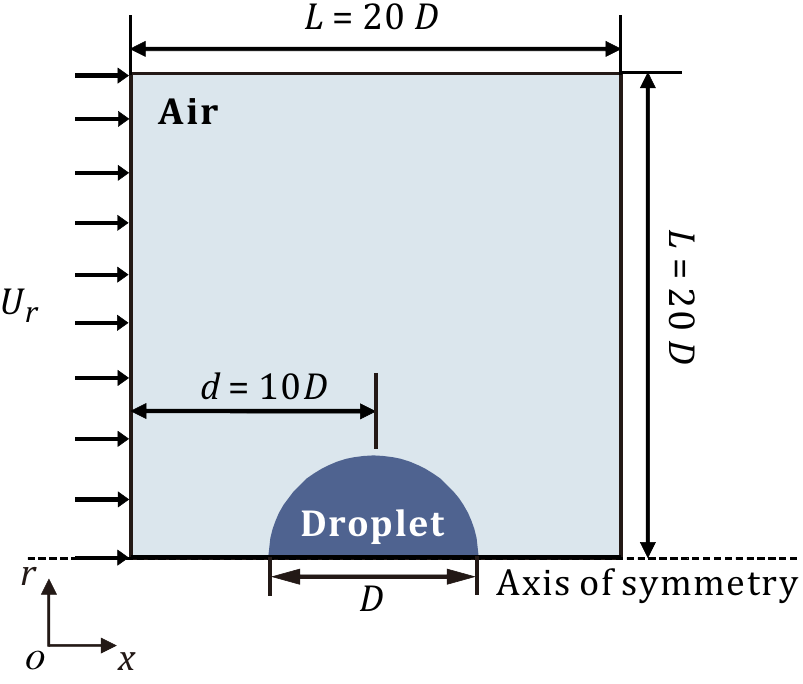}}%
  \caption{Schematic diagram for the simulation of droplet breakup in the air stream.}
\label{fig:fig01}
\end{figure}

A schematic diagram of the simulation is shown in Fig.\ \ref{fig:fig01}. An axisymmetric domain is adopted to reduce the computation cost, similar to many previous studies \cite{Cao2022KeroseneGel, Feng2010InterfacialFlows, Jain2019ModerateWeber, Marcotte2019Thresholds}. The axisymmetric setting can allow us to consider the initial spherical shape of the droplet and the axisymmetric deformation. However, once the droplet breaks up, the process is three-dimensional, and 2D axisymmetric simulation cannot fully capture the process. Therefore, with the axisymmetric setting, we mainly consider the deformation of the droplet before the breakup. Even though three-dimensional (3D) simulation is possible and adopted by some studies \cite{Yang2017Transitions, Yang2016HighlyUnstable, Chu2020PolymerSolution, Tavangar2015CoalWaterSlurry}, 2D simulation allows us to use a very fine mesh to capture the details of the flow and droplet deformation, which is important for us to unveil the effects of the shear-thinning rheology of the droplet.

The left boundary of the domain is set as uniform velocity inflow ${U_{{r}}}$. The right and the top boundaries are set free outflow, i.e., a Neumann condition for velocity ${{\partial {\bf{u}}}}/{{\partial {\bf{n}}}} = 0$ and a Dirichlet condition for pressure $p = 0$. The droplet has an initial diameter of $D=4$ mm, and it is initially located on the axis 10$D$ away from the left boundary (see the Supplementary Material for the effect of the distance from the left boundary). At the beginning of the simulation, the velocity in the whole domain is set to zero. Then the droplet gradually deforms in the air stream. To quantify the droplet deformation, we use the cross-stream diameter ${D_{{\rm{cro}}}}$ (i.e., the droplet diameter in the cross direction). This parameter is normalized by the initial droplet diameter as ${D_{{\rm{cro}}}}/D$.

\subsection{Mathematical model}\label{sec:sec022}
The system contains a droplet of a shear-thinning fluid exposed to the airflow. The droplet and air are considered as incompressible fluids with constant densities. Similar to many previous studies \cite{Cao2022KeroseneGel, Meng2018SheetInstability}, mass transfer, heat transfer, and gravity are neglected in the simulation because it is very weak in most applications of droplet deformation and breakup processes. With the VOF method, the flow field can be modeled by a single continuum \cite{Hirt1981FreeBoundaries}. The continuity equation, the momentum equation, and the transport equation of the volume fraction   \cite{Berberovic2010impact} are solved
\begin{equation}\label{eq:eq1}
  \nabla  \cdot {\bf{u}} = 0
\end{equation}
\begin{equation}\label{eq:eq2}
  \rho \left( {\frac{{\partial {\bf{u}}}}{{\partial t}} + \nabla  \cdot {\bf{uu}}} \right) =  - \nabla p + \nabla  \cdot (2\mu {\bf{D}}) + \sigma \kappa {\bf{n}}{\delta _S}\
\end{equation}
\begin{equation}\label{eq:eq3}
  \frac{{\partial f}}{{\partial t}} + \nabla  \cdot ({\bf{u}}f) = 0\
\end{equation}
where ${\bf{u}}$ is the velocity vector; $\mu $ is the dynamic viscosity; ${\bf{D}}$ is the deformation rate tensor given by ${\bf{D}} = \left[ {\nabla {\bf{u}} + {{(\nabla {\bf{u}})}^{\bf{T}}}} \right]/2$, and $\sigma $ is the surface tension coefficient. In the last item of Eq.\ (\ref{eq:eq2}), $\sigma \kappa {\bf{n}}{\delta _S}$ represents the surface tension force of the gas-liquid interface. In Eq.\ (\ref{eq:eq3}), $f $ is the volume fraction of the liquid in the control volume. It is $f $ = 1 if the control volume is full of liquid, and $f $ = 0 if full of gas. $0< f <1$ indicates the existence of an interface in the control volume.

The density and the viscosity of the fluid are expressed as:
\begin{equation}\label{eq:eq4}
  \rho  = {\rho _{{L}}}f + {\rho _{{G}}}(1 - f)
\end{equation}
\begin{equation}\label{eq:eq5}
  \mu  = {\mu _{{L}}}f + {\mu _{{G}}}(1 - f)
\end{equation}
where the subscripts $G$ and $L$ denote the gas (air) and the liquid (shear-thinning or Newtonian droplet) phases, respectively.

The power-law fluid model is adopted to describe the rheology of the shear-thinning fluid \cite{Arnold2010Gels}:
\begin{equation}\label{eq:eq6}
  {\mu _{{\rm{eff}}}} = K{\dot \gamma ^{n - 1}}
\end{equation}
where ${\mu _{{\rm{eff}}}}$ is the effective viscosity, $\dot \gamma $ is the shear rate, $K$ is the consistency index, and $n$ is the flow index of the power-law fluid. When $n = 1$, Eq.\ (\ref{eq:eq6}) describes a Newtonian fluid. When $0 < n < 1$ the fluid is shear-thinning, and when $n > 1$ the fluid is shear-thickening. In this study, the flow index $n$ and the consistency index $K$ are varied in the range of $0.4< n <1$ and $0.00569< K < 26.785$, respectively. The other properties of shear-thinning fluids, Newtonian fluids, and air used in the simulation are listed in Table \ref{tab:tab1}, except otherwise explicitly specified. The density and the surface tension of the droplet are set according to the properties of water. The viscosity of the Newtonian fluid is calculated according to the Ohnesorge number defined in Eq.\ (\ref{eq:eq11}). The governing equations are solved by the open-source multiphase flow solver Basilisk \cite{Popinet2018Basilisk}. Quadtree space discretization is adopted for adaptive mesh refinement (AMR) \cite{Lagree2011GranularMedia, Popinet2003Gerris, Popinet2009ParasiticCurrents}, which dynamically changes the grid structure according to the local interface shape and the velocity field. Hence, it can increase the simulation resolution without significantly increasing the computational demand.

To quantify the influence of shear-thinning rheological properties on the deformation and breakup process, several dimensionless numbers are used, including the Ohnesorge number, the Weber number, and the dimensionless time. The Ohnesorge number $Oh$ and the Weber number $\We$ can be obtained by nondimensionalize the momentum equation (\ref{eq:eq2}).
By introducing the following variables, ${\bf{x}}^* = {\bf{x}}/D$, ${\bf{u}}^* = {\bf{u}}/{U_{{r}}}$, $t^* = t{U_{{r}}}/D$, $p^* = p/({\rho _{{L}}}{U_{{r}}}^2)$, $\kappa^* = D\kappa$, ${\delta^*_S} = D{\delta _S}$, $\mu^* = \mu/{\mu _{\text{eff}}}$, $\rho^* = \rho/{\rho _{{L}}}$, the momentum equation can be nondimensionalized,
\begin{equation}\label{eq:eq2.6}
\begin{aligned}
  &{\frac{{\partial {\bf{u}}^*}}{{\partial t^*}} + {\bf{u}}^* \cdot \nabla  {\bf{u}}^*}   =  - \nabla p^* \\
  + & \frac{Oh}{\sqrt{\We}} \nabla  \cdot \mu^* \left( {\nabla {\bf{u}}^* + {{\nabla {\bf{u}}^*}^{\bf{T}}}} \right) + \frac{1}{\We} \kappa^* {\bf{n}}{\delta_S^*}.
\end{aligned}
\end{equation}
It should be noted that in the definition of the Ohnesorge number, the viscosity of the fluid is involved, which is not a constant for the shear-thinning droplet and depends on the shear rate. Hence, the Ohnesorge number $Oh$ in Eq.\ (\ref{eq:eq2.6}) is defined in Eq.\ (\ref{eq:eq11}), which can consider the characteristic effective viscosity at the characteristic shear rate in the shear thinning droplet, and the detail will be discussed in Sec.\ \ref{sec:sec031}. The Weber number $\We$ in Eq.\ (\ref{eq:eq2.6}) represents the ratio of inertia to surface tension force,
\begin{equation}\label{eq:eq7}
  \We = \frac{{{\rho _{{G}}}{U_{{r}}}^2D}}{\sigma }.
\end{equation}
The Weber number in the simulations is varied by changing the initial relative velocity between gas and liquid phases (${U_{{r}}}$).
Instead of using $t^*$ directly, the dimensionless time is defined following Nicholls and Ranger \cite{Nicholls1969AerodynamicShattering}:
\begin{equation}\label{eq:eq8}
  T = \sqrt {\frac{{{\rho _{{G}}}}}{{{\rho _{{L}}}}}} \frac{{{U_{{r}}}}}{D}t.
\end{equation}
Hence, $T= \sqrt{{\rho_G}/{\rho_L}} t^*$.

\begin{table*}[]
\setlength{\tabcolsep}{8pt}
\caption{Physical properties of the fluid used in the simulation.}
\begin{tabular}{ccccl}
\hline
Fluids               & Viscosity, $\mu$ (Pa$\cdot$s) & Density, $\rho$ (kg/m$^3$) & Surface tension, $\sigma$ (N/m)\\
\hline
Shear-thinning fluid & Eq.\ (\ref{eq:eq6})            & 1000            & 0.07                     \\
Newtonian fluid      & $Oh{{\sqrt{{\rho _{{L}}}D\sigma}}}$  & 1000            & 0.07                    \\
Air                  & $1.8\times{10^{-5}}$           & 1.2             & -                        \\
\hline
\end{tabular}
\label{tab:tab1}
\end{table*}

\begin{figure}
  \centerline{\includegraphics[width=0.8\columnwidth]{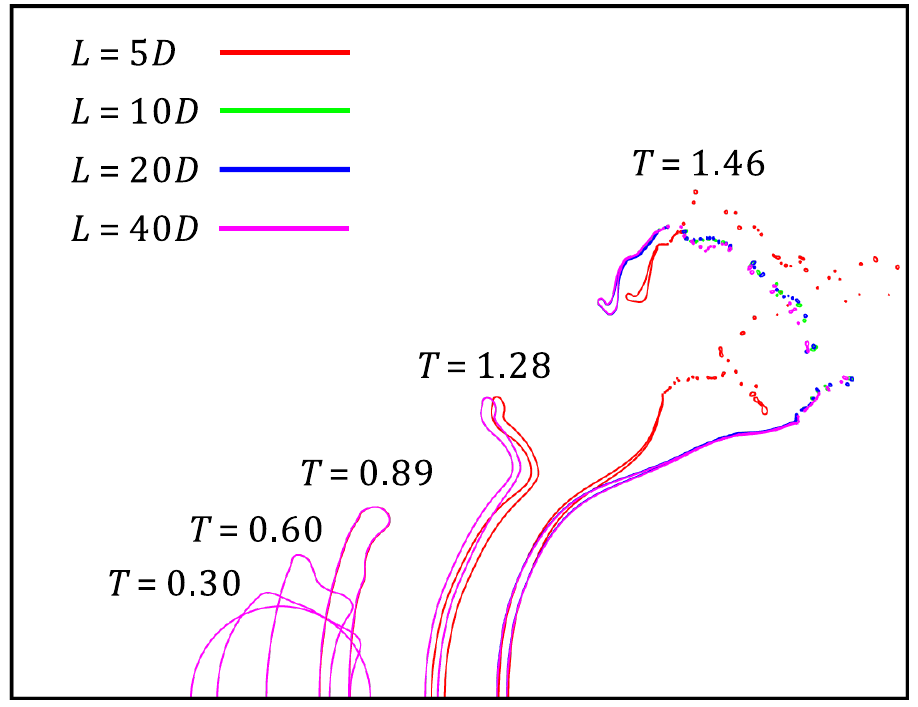}}%
  \caption{Evolutions of droplet morphology for various domain sizes ($L\times L$).}
\label{fig:fig02}
\end{figure}

In the simulation, the size of the computational domain needs to be large enough that the selection of the boundary does not affect the simulation results. To select the proper size of the computational domain, four sizes were tested and the results are shown in Fig.\ \ref{fig:fig02}. The comparison shows that the difference in the early stage of droplet deformation is indistinguishable. As the droplet deforms, the differences gradually become noticeable ($T = 1.28$). The droplet morphology after the breakup ($T=1.46$) shows that the domain size of $L = 20 D$ is sufficient to capture the whole process of droplet deformation and breakup, therefore, $L = 20 D$ is selected as the default domain size for further simulations.

\begin{figure}
  \centerline{\includegraphics[width=0.75\columnwidth]{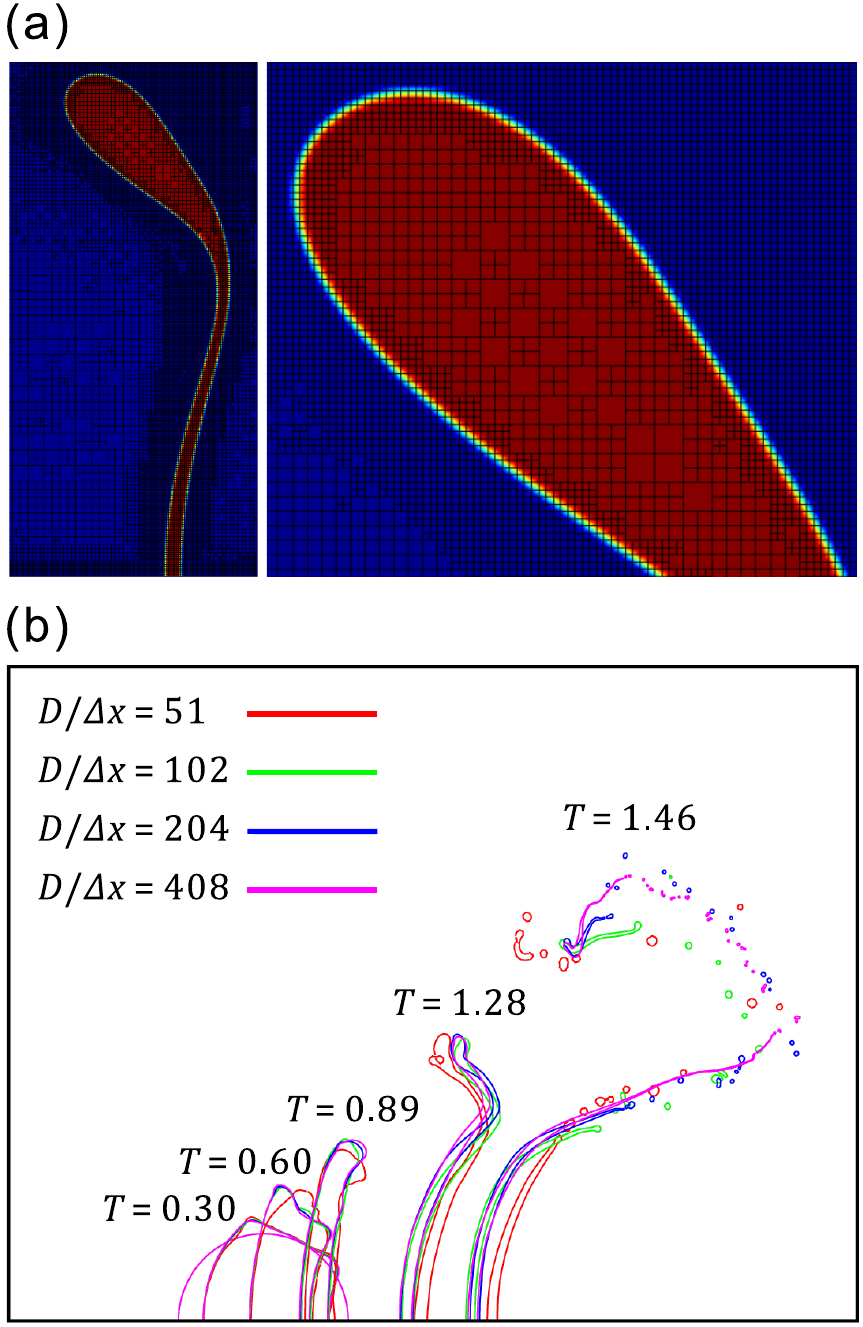}}
  \caption{(a) Adaptive grid refinement used in this study. (b) Droplet shapes at different grid resolutions. Here, $\We$ = 20, $Oh$ = 0.008 and $n$ = 0.5 for the shear-thinning droplet. The air flows from left to right.}
\label{fig:fig03}
\end{figure}

The mesh size in the simulation determines the accuracy and numerical stability, which is crucial for capturing the liquid-gas interface during the droplet deformation and breakup. The adaptive mesh refinement is performed as shown in Fig.\ \ref{fig:fig03}(a). Compared with static global mesh refinement, the computation time and storage cost of adaptive mesh refinement can be greatly reduced. We progressively increase the maximum level of refinement and compare the results. Fig.\ \ref{fig:fig03}(b) shows the contours of droplets at four mesh resolutions of $D/\Delta x$ = 51, 102, 204, and 408, where $\Delta x$ is the minimum mesh size. The results show that when $D/\Delta x$ is greater than 204, the droplet deformation and breakup are almost independent of the mesh size. Therefore, $D/\Delta x = 204$ is used in the following simulations to study the deformation and breakup of droplets.

\subsection{Experimental validation}\label{sec:sec023}
\begin{figure*}
  \centerline{\includegraphics[width=1.8\columnwidth]{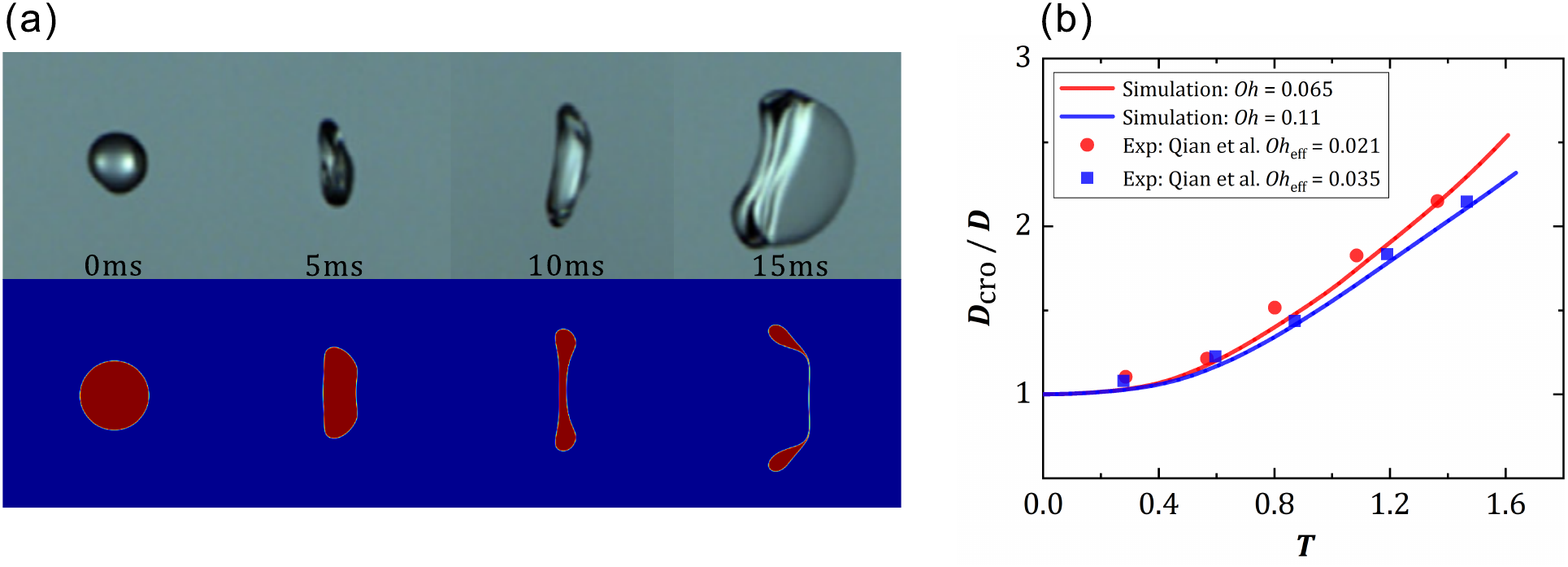}}
  \caption{Comparison of the numerical simulation against the experimental results in Ref.\ \citenum{Qian2021CarboxymethylCelluloseDroplets}. (a) Bag breakup of the shear-thinning droplet with $\We$ = 16. The time interval is 5 ms. The first row is the experimental images in Ref.\ \citenum{Qian2021CarboxymethylCelluloseDroplets} and the second row is the numerical results of this study. The experimental images are reprinted from International Journal of Multiphase Flow, 136, Qian et al., ``An experimental investigation on the secondary breakup of carboxymethyl cellulose droplets'', 103526, Copyright (2021), with permission from Elsevier. (b) Evolution of cross-stream diameter (${{D}_{\rm{cro}}}/D$) in the multi-bag breakup mode with $\We$ = 33. It should be noted that the Ohnesorge number is defined differently in Ref.\ \citenum{Qian2021CarboxymethylCelluloseDroplets}, $O{{h}_{\rm{eff}}}={K}/{\sqrt{{{\rho }_{{L}}}\sigma {{D}^{2n-1}}U_{{r}}^{2-2n}}}$, hence its relation to the definition in Eq.\ (\ref{eq:eq11}) is $Oh={O{{h}_{\rm{eff}}}}/{\sqrt{{{({{\rho }_{{G}}}/{{\rho }_{{L}}})}^{1-n}}}}$}
\label{fig:fig04}
\end{figure*}

The numerical results are validated against experimental data in Ref.\ \citenum{Qian2021CarboxymethylCelluloseDroplets}, as shown in Fig.\ \ref{fig:fig04}. The droplet used in Ref.\ \citenum{Qian2021CarboxymethylCelluloseDroplets} is a 0.2\% CMC solution, which is a shear-thinning fluid. The deformation and breakup process of a droplet with $\We = 16$ from the experiment and simulation is compared in Fig.\ \ref{fig:fig04}(a). The numerical result is consistent with the experimental result and can reproduce the bag breakup process. In addition, for the multimode breakup, the evolution of the cross-stream diameter ${D_{{\rm{cro}}}}/D$ of two droplets ($Oh = 0.065$ and 0.11) at $\We$ = 33 from the experiment and simulations is shown in Fig.\ \ref{fig:fig04}(b). It can be seen that this numerical model can capture the morphological changes of the droplets reasonably well.

\section{Results and discussions}\label{sec:sec03}
\subsection{Comparison between Newtonian and shear-thinning droplets}\label{sec:sec031}
\begin{figure*}
  \centerline{\includegraphics[width=1.7\columnwidth]{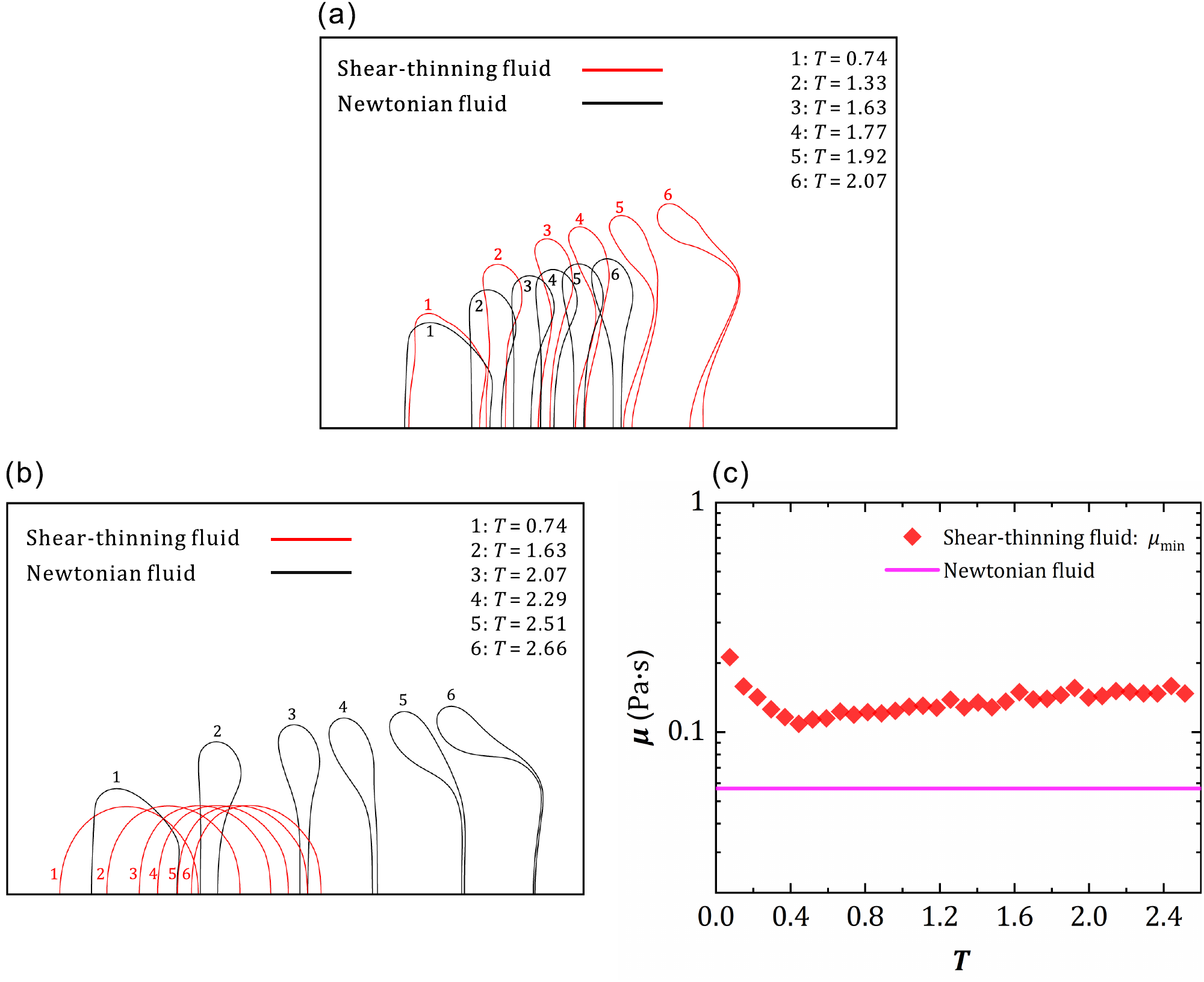}}
  \caption{Comparison of the droplet shape in the different breakup regimes between the shear-thinning and Newtonian droplets. Droplet morphology during the breakup process: (a) $n = 0.4$, $K = 0.056862$ and (b) $n = 0.4$, $Oh$ = 0.10746 of the shear-thinning droplet. (c) Minimum effective viscosity ${{\mu }_{\min }}$ inside the shear-thinning droplet. Here, $\We$ = 20, and $Oh$ = 0.10746 of the Newtonian droplet.}
\label{fig:fig05}
\end{figure*}

To compare the breakup between Newtonian and shear-thinning droplets, two approaches are widely used. One is to simply change the flow index $n$. If $n = 1$ the fluid is Newtonian, and if $n < 1$ the fluid is shear-thinning. This approach has been widely used in many studies \cite{Cao2022KeroseneGel, Kant2022BreakupMechanism, Qian2021CarboxymethylCelluloseDroplets}. The breakup process of two droplets using this approach is compared in Fig.\ \ref{fig:fig05}(a). We can see that as the fluid changes from the Newtonian to shear-thinning ($n = 0.4$), there is a significant change in the droplet breakup process. The two breakup processes are in different modes at the same $\We$, i.e., the Newtonian droplet is in the bag breakup regime while the shear-thinning droplet is in the multimode breakup regime. This is because in this approach, the magnitudes of viscosities of the two droplets are significantly different. When comparing Newtonian and shear-thinning fluids, it is important to have the same characteristic viscosity. To keep the relative importance of the viscosity the same, a second approach is widely used. Because of the shear-thinning property, the effective viscosity in the droplet is different everywhere. Therefore, a characteristic viscosity should be selected, corresponding to a characteristic shear rate. Many studies used ${U_{{r}}}/D$ as the characteristic shear rate \cite{Chu2020PolymerSolution, Kant2022BreakupMechanism}. By using this, we can maintain the same effective viscosity by adjusting $K$ when we change $n$. The comparison between a Newtonian droplet and a shear-thinning droplet by this approach is shown in Fig.\ \ref{fig:fig05}(b). We can see that the two breakup processes are also in different modes, i.e., the Newtonian droplet is in the bag breakup regime while the shear-thinning droplet is in the vibrational regime. Moreover, the minimum effective viscosity ${\mu _{\min }}$ of the shear-thinning droplet is actually always larger than that of the Newtonian droplet as shown in Fig.\ \ref{fig:fig05}(c). This indicates that the shear-thinning droplet is too more viscous than the Newtonian droplet, and it is not reasonable to compare these two cases by this approach.

To make a reasonable comparison between Newtonian and shear-thinning droplets, we consider this process by examining the characteristic shear rate. For droplet breakup, the characteristic velocity that determines the characteristic shear rate should be the characteristic velocity in the droplet ${U_{{L}}}$, not the characteristic velocity in the air ${U_{{G}}}$. Therefore, the characteristic shear rate ($\dot \gamma $) should be $\dot \gamma  = {U_{{L}}}/D$. Since the flow velocity in the droplet is induced by the external airflow, we can estimate the characteristic internal velocity of the droplet by considering the inertia of the air and the droplet fluid. Compared with other forces (such as the viscous force), the kinetic energy of the air is the main contribution to the deformation and the breakup of the droplet. Therefore, the inertia of the air should be at the same scale as the inertia of the droplet fluid, ${\rho _{{G}}}{U_{{G}}}^2/2 \sim {\rho _{{L}}}{U_{{L}}}^2/2$, where ${U_{{G}}}$ and ${U_{{L}}}$ are the characteristic velocities in the air and the liquid. As the airflow velocity ${U_{{G}}}$ can be characterized by the relative airflow velocity ${U_{{r}}}$, we can then have ${U_L} = {U_{{r}}}\sqrt {{{{\rho _{{G}}}}}/{{{\rho _{{L}}}}}} $. Therefore, the characteristic shear rate can be estimated as
\begin{equation}\label{eq:eq9}
  \dot \gamma  = \sqrt {\frac{{{\rho _{{G}}}}}{{{\rho _{{L}}}}}} \frac{{{U_{{r}}}}}{D}
\end{equation}
Then, the characteristic effective viscosity of the droplets in the deformation and breakup process can be written as
\begin{equation}\label{eq:eq10}
  {\mu _{{\rm{eff}}}} = K{\left( {\sqrt {\frac{{{\rho _{{G}}}}}{{{\rho _{{L}}}}}} \frac{{{U_{{r}}}}}{D}} \right)^{n - 1}}
\end{equation}
Hence, the Ohnesorge number can be expressed as:
\begin{equation}\label{eq:eq11}
  Oh = \frac{{{\mu _{{\rm{eff}}}}}}{{\sqrt {{\rho _{{L}}}D\sigma } }} = \frac{K}{{\sqrt {{\rho _{{L}}}\sigma {D^{2n - 1}}U_{{r}}^{2 - 2n}{{({\rho _{{G}}}/{\rho _{{L}}})}^{1 - n}}} }}
\end{equation}
For Newtonian fluids (i.e., $n = 1$ and $K = {\mu _{{L}}}$), the Ohnesorge number recovers the original definition. Since both $K$ and $n$ affect the Ohnesorge number, in some simulations, to maintain the same $Oh$ while changing $n$ or ${U_{{r}}}$, we change $K$ accordingly.

\begin{figure*}
  \centerline{\includegraphics[width=1.8\columnwidth]{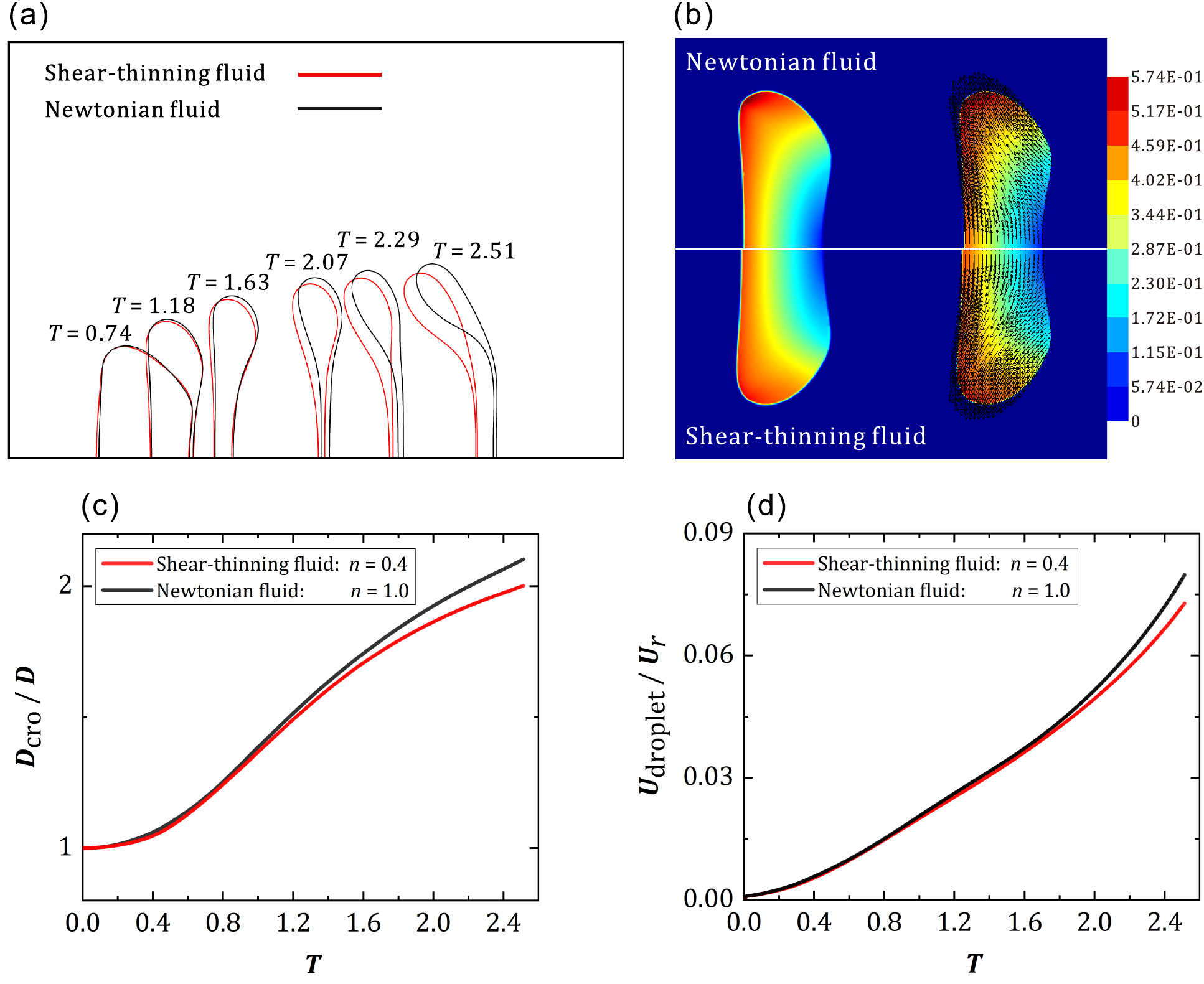}}
  \caption{Comparison of the droplet shape in the bag breakup regime between the shear-thinning and Newtonian droplets. (a) Droplet morphology during the breakup process (Multimedia view). (b) Internal velocity field and velocity vectors at $T$ = 1.04. The air flows from left to right. (c) Evolution of the cross-stream diameter ${{D}_{\rm{cro}}}/D$. (d) Evolution of the droplet centroid velocity ${{U}_{\rm{droplet}}}/{{U}_{{r}}}$. Here, $\We$ = 20 and $Oh$ = 0.10746, and $n$ = 0.4 for the shear-thinning droplet.}
\label{fig:fig06}
\end{figure*}

Based on the above selection of the characteristic shear rate and the definition of the Ohnesorge number, the deformation and breakup processes of Newtonian droplets and shear-thinning droplets are considered under the same $\We$ number and same $Oh$ number, i.e., the same characteristic shear rate and the same effective viscosity. Two typical cases are considered first, one in the bag breakup regime ($\We  = 20$, $Oh  = 0.10746$, Fig.\ \ref{fig:fig06}), and the other in multimode breakup regime ($\We  = 30$, $Oh= 0.010746$, Fig.\ \ref{fig:fig08}). The comparison can show that, using this proposed approach, the processes for Newtonian and shear-thinning droplets are comparable. Therefore, the current selection of the characteristic shear rate and the definition of the Ohnesorge number is better than the previous approaches.

For the bag breakup ($\We  = 20$, $Oh  = 0.10746$), in the early stage, the difference in the deformation between the two droplets is small as shown in Fig.\ \ref{fig:fig06}(a) (Multimedia view). Comparing the velocity fields between the two droplets, we can see that the Newtonian droplet has a high-velocity region on the windward side near the rim, and the internal velocity field is more uniform for the shear-thinning droplets as shown in Fig.\ \ref{fig:fig06}(b). This will lead to differences in the later deformation of the droplet. In addition, the change rates of ${D_{{\rm{cro}}}}/D$, and ${U_{{\rm{droplet}}}}/{U_{{r}}}$ in these two droplets are almost the same as shown in Fig.\ \ref{fig:fig06}. While in the later stage, compared with Newtonian fluids, the deformation of the shear-thinning droplet near the center increases, and the center of droplets becomes thinner. The change rate of the cross-stream diameter and the centroid velocity of the shear-thinning droplet is also slower than that of the Newtonian droplet, as shown in Figs.\ \ref{fig:fig06}(c) and \ref{fig:fig06}(d).

\begin{figure*}
  \centerline{\includegraphics[width=2\columnwidth]{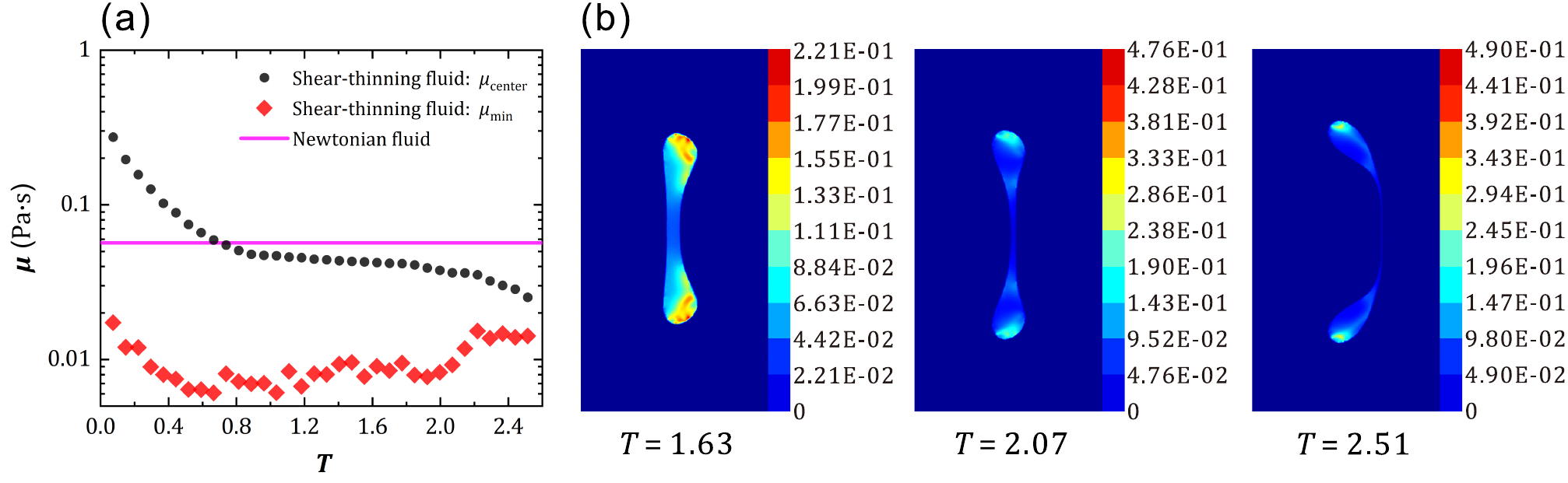}}
  \caption{Evolution of the effective viscosity in the shear-thinning droplet: (a) The effective viscosity at the droplet center ${{\mu }_{\rm{center}}}$ and the minimum effective viscosity of the shear-thinning droplet ${{\mu }_{\min }}$. (b) Effective viscosity distribution inside the shear-thinning droplet. Here, $\We$ = 20 and $Oh$ = 0.10746, and $n$ = 0.4.}
\label{fig:fig07}
\end{figure*}

\begin{figure*}
  \centerline{\includegraphics[width=1.8\columnwidth]{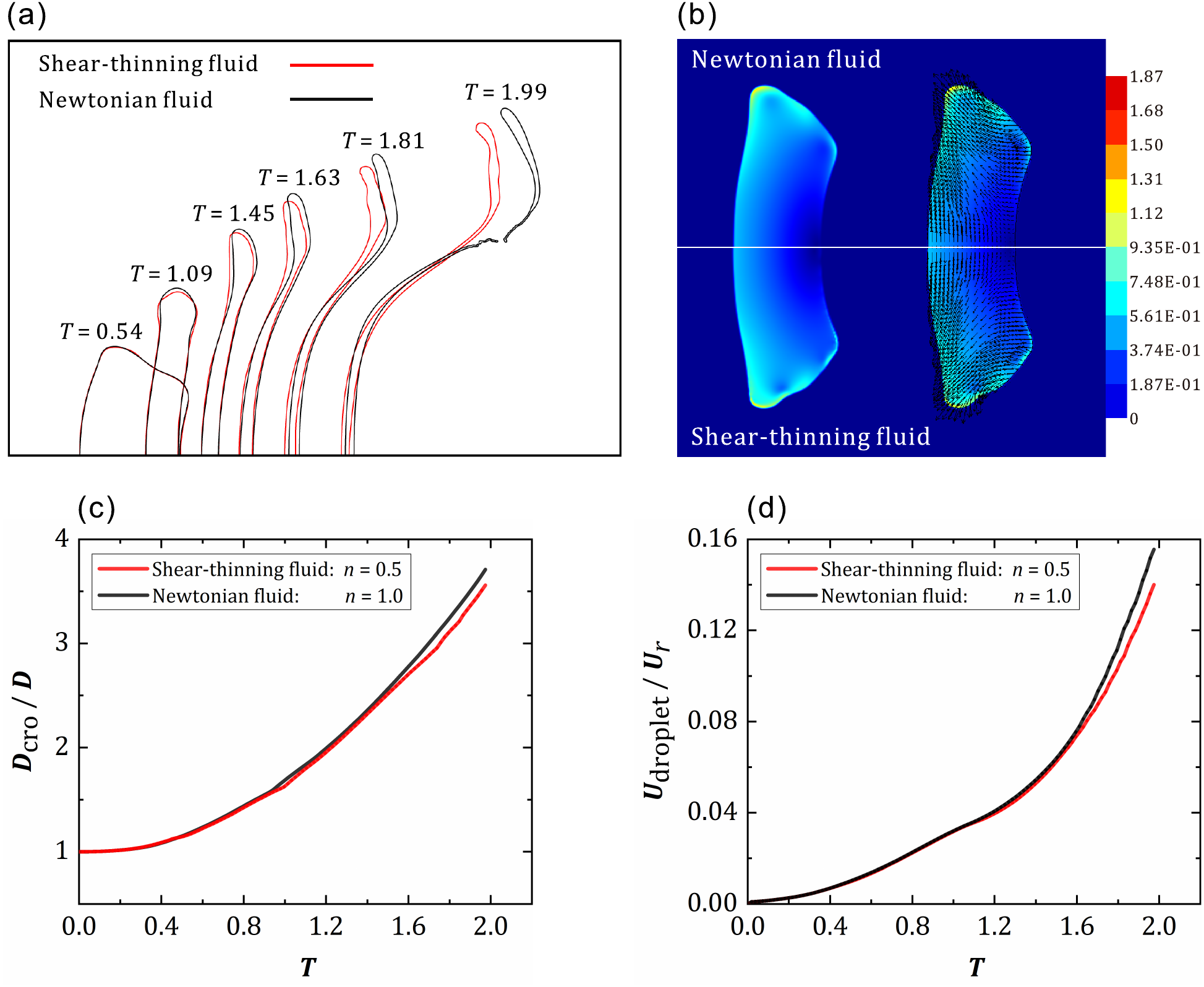}}
  \caption{Comparison of the droplet shape in the multimode breakup regime between the shear-thinning and Newtonian droplets. (a) Droplet morphology during the breakup process (Multimedia view). (b) Internal velocity field and velocity vectors at $T$ = 0.80. The air flows from left to right. (c) Evolution of the cross-stream diameter ${{D}_{\rm{cro}}}/D$. (d) Evolution of the droplet centroid velocity ${{U}_{\rm{droplet}}}/{{U}_{{r}}}$. Here, $\We$ = 30 and $Oh$ = 0.010746, and $n$ = 0.5 for the shear-thinning droplet.}
\label{fig:fig08}
\end{figure*}

To analyze the origin of the difference between the Newtonian droplet and the shear-thinning droplet in the bag breakup process, the effective viscosity at the droplet center (i.e., the center point on the axis) ${\mu _{{\rm{center}}}}$ and the minimum effective viscosity ${\mu _{\min }}$ of the shear-thinning droplet are shown in Fig.\ \ref{fig:fig07}(a). In the later stage, due to the shear-thinning property, ${\mu _{{\rm{center}}}}$ and ${\mu _{\min }}$ are much smaller than the viscosity of the Newtonian droplet. The distribution of the effective viscosity in the shear-thinning droplet is shown in Fig.\ \ref{fig:fig07}(b). Even though the characteristic viscosity of the shear-thinning droplet (the effective viscosity at the characteristic shear rate $\dot \gamma $ in Eq.\ (\ref{eq:eq9})) is the same as that of the Newtonian droplet, the effective viscosity is varied within the shear-thinning droplet, which affects the deformation of the droplet. There is a low-viscosity region near the center of the shear-thinning droplet, which enhances the deformation of the droplet center. The position of the minimum effective viscosity is mainly near the center of droplets as shown in Fig.\ \ref{fig:fig07}(b). Compared with the Newtonian droplet, the effective viscosity around the center of the shear-thinning droplet decreases, and the deformation of the droplet increases.

For the multimode breakup ($\We  = 30$, $Oh= 0.010746$), the difference in the droplet shape between the Newtonian droplet and the shear-thinning droplet in the early stage is also small, as shown in Fig.\ \ref{fig:fig08}(a) (Multimedia view). By comparing the velocity fields of the two droplets, we can see that the high-speed region is distributed near the rim of the shear-thinning droplet, while it is mainly on the windward side of the rim of the Newtonian fluid, as shown in Fig.\ \ref{fig:fig08}(b). Such difference will lead to different extent of deformation near the rim of the droplets. In addition, this can also be shown quantitatively by ${D_{{\rm{cro}}}}/D$, and ${U_{{\rm{droplet}}}}/{U_{{r}}}$ in Figs.\ \ref{fig:fig08}(c) and \ref{fig:fig08}(d). In the later deformation stage, the radial stretching of the shear-thinning droplet near the rim is stronger than that of the Newtonian droplet, while the Newtonian droplet shows stronger forward stretching near the rim and higher overall acceleration. Hence, the change rates of ${{D}_{\rm{cro}}}/D$ and ${{U}_{\rm{droplet}}}/{{U}_{{r}}}$ of the Newtonian droplet are larger than that of the shear-thinning droplet.

\begin{figure*}
  \centerline{\includegraphics[width=2\columnwidth]{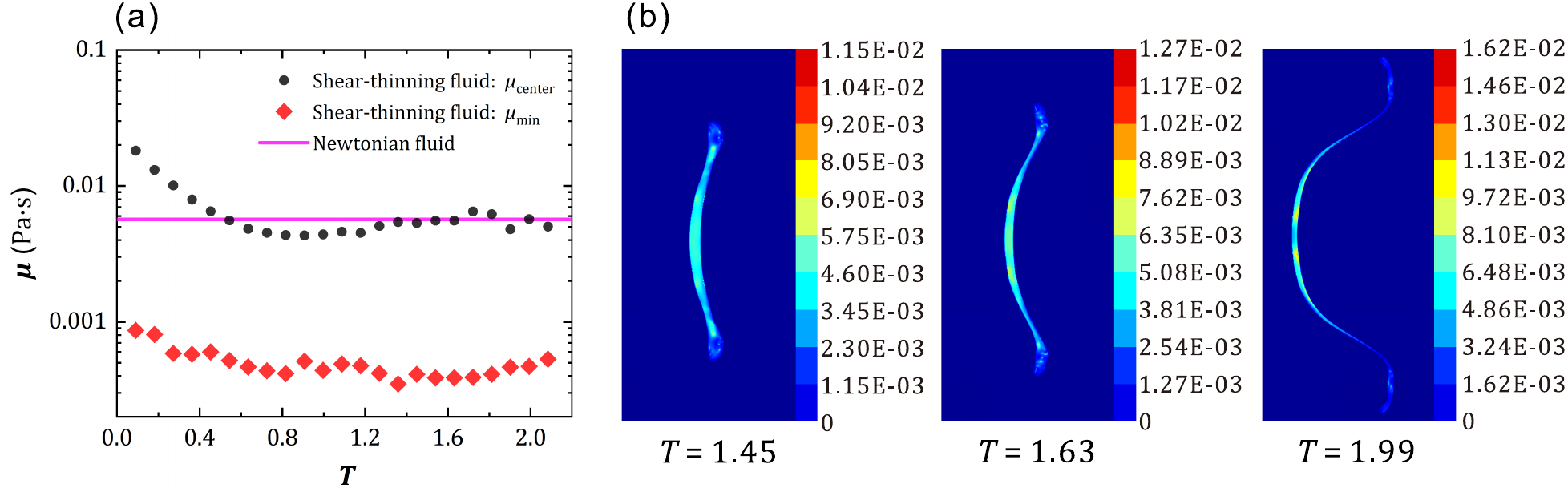}}
  \caption{Evolution of the effective viscosity in the shear-thinning droplet. (a) The effective viscosity at the droplet center ${{\mu }_{\rm{center}}}$ and the minimum effective viscosity of the shear-thinning droplet ${{\mu }_{\min }}$. (b) Effective viscosity distribution inside the shear-thinning droplet. Here, $\We$ = 30 and $Oh$ = 0.010746, and $n$ = 0.5.}
\label{fig:fig09}
\end{figure*}

To explain the differences between the two fluids in the multimode breakup process, the effective viscosity at the droplet center ${{\mu }_{\rm{center}}}$ and the minimum effective viscosity ${{\mu }_{\min }}$ of the shear-thinning droplet are shown in Fig.\ \ref{fig:fig09}(a). The minimum effective viscosity ${{\mu }_{\min }}$ inside the shear-thinning droplet is much smaller than that of the Newtonian fluid and ${{\mu }_{\rm{center}}}$ is close to the viscosity of the Newtonian droplet. In addition, the distribution of the local viscosity within the shear-thinning droplet is shown in Fig.\ \ref{fig:fig09}(b), which shows that the viscosity distribution in the shear-thinning droplet is very different from the Newtonian droplet. In the later stage, due to the shear-thinning property, the local viscosity in the outer region is much smaller than that in the inner region, as shown in Fig.\ \ref{fig:fig09}(b). Compared with the bag breakup case, the effective viscosity is smaller and the aerodynamic effect is stronger in this case. There is also a low-viscosity region in this case but it has transferred from the center to the rim of the shear-thinning droplet, as shown in Fig.\ \ref{fig:fig09}(b). The low-viscosity region near the rim of the shear-thinning droplet enhances the deformation of the droplet rim. Therefore, the shear-thinning droplet has a strong radial stretching around the rim than the Newtonian droplet.

\subsection{Effect of flow index \boldmath{$n$}}\label{sec:sec032}
\begin{figure*}
  \centerline{\includegraphics[width=2\columnwidth]{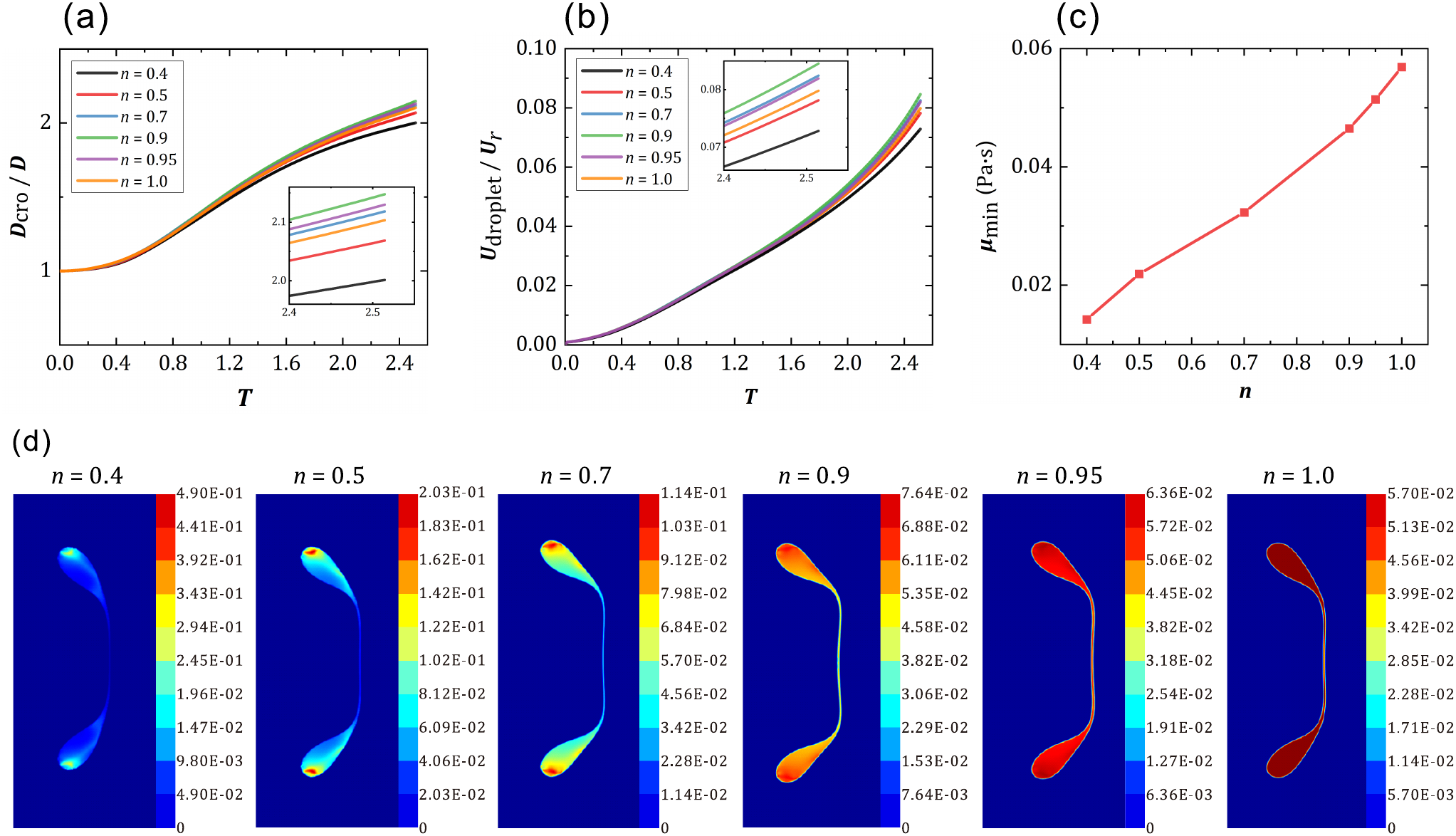}}
  \caption{Evolution of (a) the cross-stream diameter ${{D}_{\rm{cro}}}/D$ and (b) the centroid velocity ${{U}_{\rm{droplet}}}/{{U}_{{r}}}$ of droplets. (c) Minimum effective viscosity ${{\mu }_{\min }}$, and (d) viscosity distribution inside the shear-thinning droplet for various $n$ values at $T$ = 2.51. Here, $\We$ = 20 and $Oh$ = 0.10746.}
\label{fig:fig10}
\end{figure*}

The influence of the rheological parameters on the deformation and breakup mode in the process of droplet breakup is analyzed by changing $n$. The value of $Oh$ is fixed to ensure that the effective viscosity in the simulation is comparable. The effect of $n$ on the droplet deformation in the bag breakup mode is analyzed by decreasing $n$ from 1 to 0.4, while keeping $\We = 20$ and $Oh = 0.10746$. As the value of $n$ decreases, the change rates of ${{D}_{\rm{cro}}}/D$ and ${{U}_{\rm{droplet}}}/{{U}_{{r}}}$ first increase slightly and then decrease, and the maximum change rates of these two parameters are at $n = 0.9$ as shown in Figs.\ \ref{fig:fig10}(a) and \ref{fig:fig10}(b).

This effect can be explained by the shear-thinning property of the droplet. The smaller the value of $n$ is, the stronger the shear-thinning effect is. Hence, as $n$ decreases, the local effective viscosity inside droplets has a significant change, and the minimum effective viscosity ${{\mu }_{\min }}$ decreases as shown in Fig.\ \ref{fig:fig10}(c). In addition, the difference in the viscosity within the droplet is more obvious as shown in Fig.\ \ref{fig:fig10}(d). Because the low-viscosity region is mainly located near the droplet center, the deformation near the droplet center is gradually enhanced. In contrast, the high viscosity around the rim of the droplet hinders the deformation of the droplet. In addition, when $n$ approaches 1, the viscosity distribution is almost uniform, and there is no easy-to-deform region. Hence, without strong local deformation, the deformation of the droplet is also relatively slow when $n$ approaches 1. Therefore, the shear-thinning property causes the transition of the change rates of ${{D}_{\rm{cro}}}/D$ and ${{U}_{\rm{droplet}}}/{{U}_{{r}}}$.

\begin{figure*}
  \centerline{\includegraphics[width=2\columnwidth]{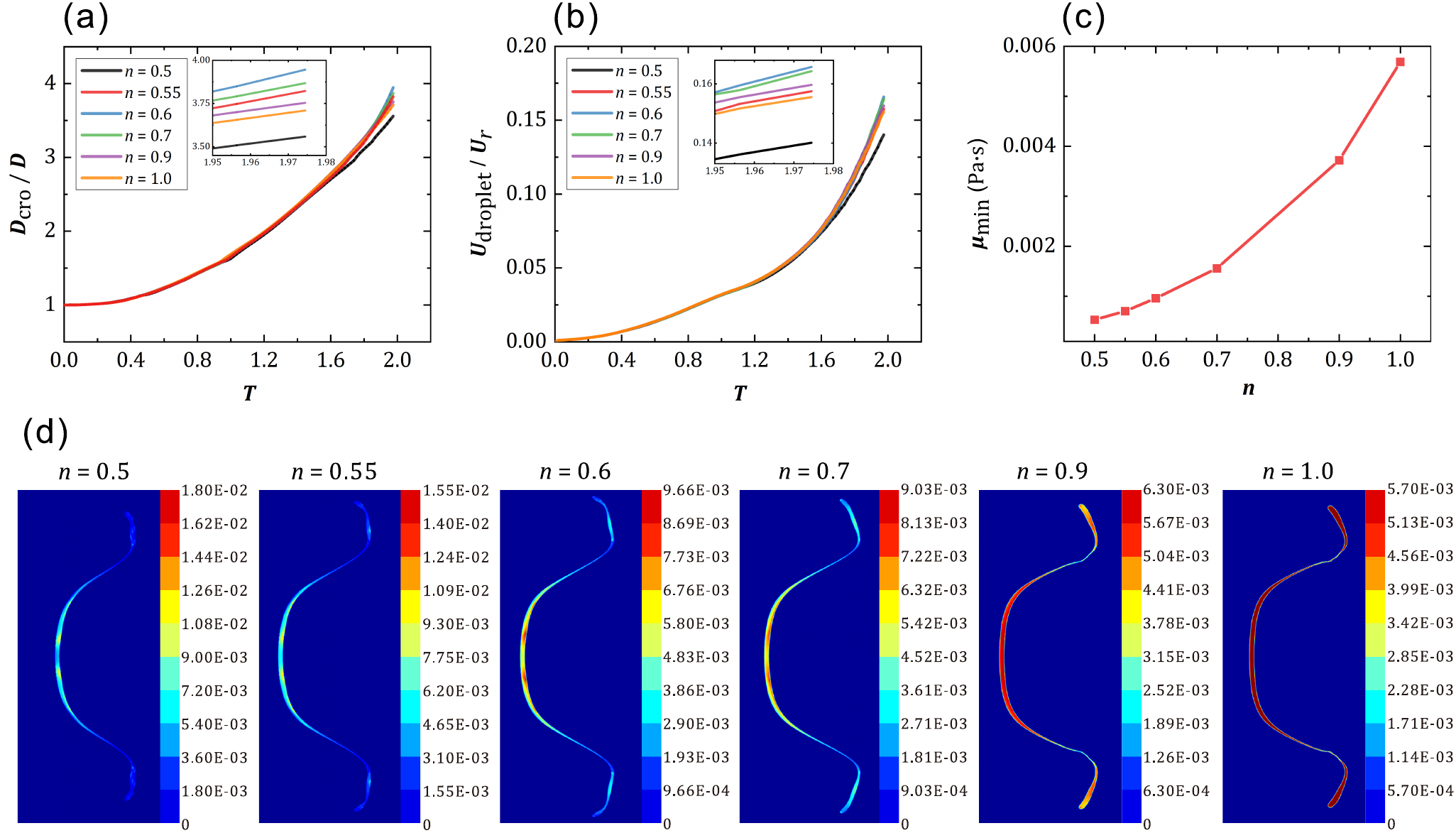}}
  \caption{Evolution of (a) the cross-stream diameter ${{D}_{\rm{cro}}}/D$ and (b) the centroid velocity ${{U}_{\rm{droplet}}}/{{U}_{{r}}}$ of droplets. (c) Minimum effective viscosity ${{\mu }_{\min }}$ and (d) viscosity distribution inside the shear-thinning droplet for various $n$ values at $T$ = 1.97. Here, $\We$ = 30 and $Oh$ = 0.010746.}
\label{fig:fig11}
\end{figure*}

The effect of $n$ on the multimode breakup process ($\We$ = 30, $Oh$ = 0.010746) is also analyzed. As $n$ increases, the change rates of ${{D}_{\rm{cro}}}/D$ and ${{U}_{\rm{droplet}}}/{{U}_{{r}}}$ also first increase and then decrease, and the maximum change rates of these two parameters are at $n = 0.6$ as shown in Figs.\ \ref{fig:fig11}(a) and \ref{fig:fig11}(b). For this case, as $n$ decreases, the shear-thinning effect increases, and the minimum effective viscosity ${{\mu }_{\min }}$ also decreases as shown in Fig.\ \ref{fig:fig11}(c). In addition, the low-viscosity region is mainly located near the droplet rim and the differences between high-viscosity and low-viscosity regions are significant as shown in Fig.\ \ref{fig:fig11}(d). As $n$ decreases, the radial stretching of the droplet at the rim is gradually enhanced. While the high viscosity region near the center of droplets also hinders the deformation. In addition, when $n$ approaches 1, the deformation is also hindered because the viscosity distribution is almost uniform and there is no easy-to-deform region. Therefore, as $n$ decreases, there is a change for the rim, i.e., from the forward stretching to the radial stretching. This effect can also be explained by the shear-thinning property. When $n$ decreases, the local effective viscosity decreases due to high local velocity and high local shear rate. Therefore, with the decrease of the local effective viscosity, the rim of the droplet is more likely to be accelerated and make stronger radial stretching at a smaller $n$ value.

\subsection{Effect of Weber number}\label{sec:sec033}
\begin{figure}
  \centerline{\includegraphics[width=0.8\columnwidth]{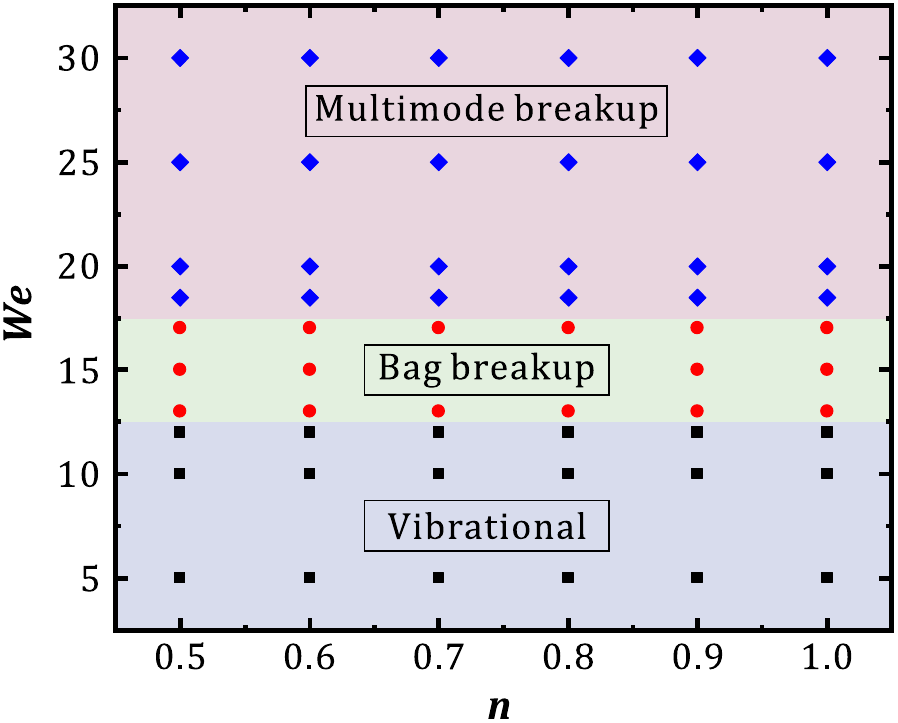}}
  \caption{Regime map of droplet breakup in the $\We$--$n$ space. Here, $Oh$ = 0.010746.}
\label{fig:fig12}
\end{figure}

The effect of the Weber number $\We$ on the deformation and breakup mode of shear-thinning droplets is analyzed in this section. To consider the effect of $\We$, we fixed other parameters ($Oh$ and $n$) and changed the airflow velocity ${{U}_{{r}}}$ from 8.54 to 22.59 m/s (that is, $\We$ from 5 to 35) to explore its influence. Meanwhile, to maintain the constant $Oh$ as ${{U}_{{r}}}$ changes, we change $K$ accordingly. Then, we change the Weber number $\We$ and the power-law index $n$, and produce a regime map for the breakup mode in the $\We$--$n$ space, as shown in Fig.\ \ref{fig:fig12}. We can see that the increase of $\We$ alters the breakup mode: as the $\We$ increases, the droplet changes from the vibrational to the bag breakup, and finally to the multimode breakup. This could be attributed to the higher inertia at large $\We$, which accelerates the droplet deformation remarkably. In addition, the results show that $n$ does not influence the breakup mode. This is mainly because we have considered the effect of $n$ on the viscosity by using a modified definition of the Ohnesorge number by Eq.\ (\ref{eq:eq11}). Hence, when we change $n$, the overall effect of the viscosity does not change because $Oh$ is fixed at a low value and the effect of $n$ is mainly on the local viscosity distribution and the local deformation of the droplet. We will see that the change of $n$ does affect the breakup mode when the $Oh$ number is large in Section \ref{sec:sec034}.

\begin{figure*}
  \centerline{\includegraphics[width=2\columnwidth]{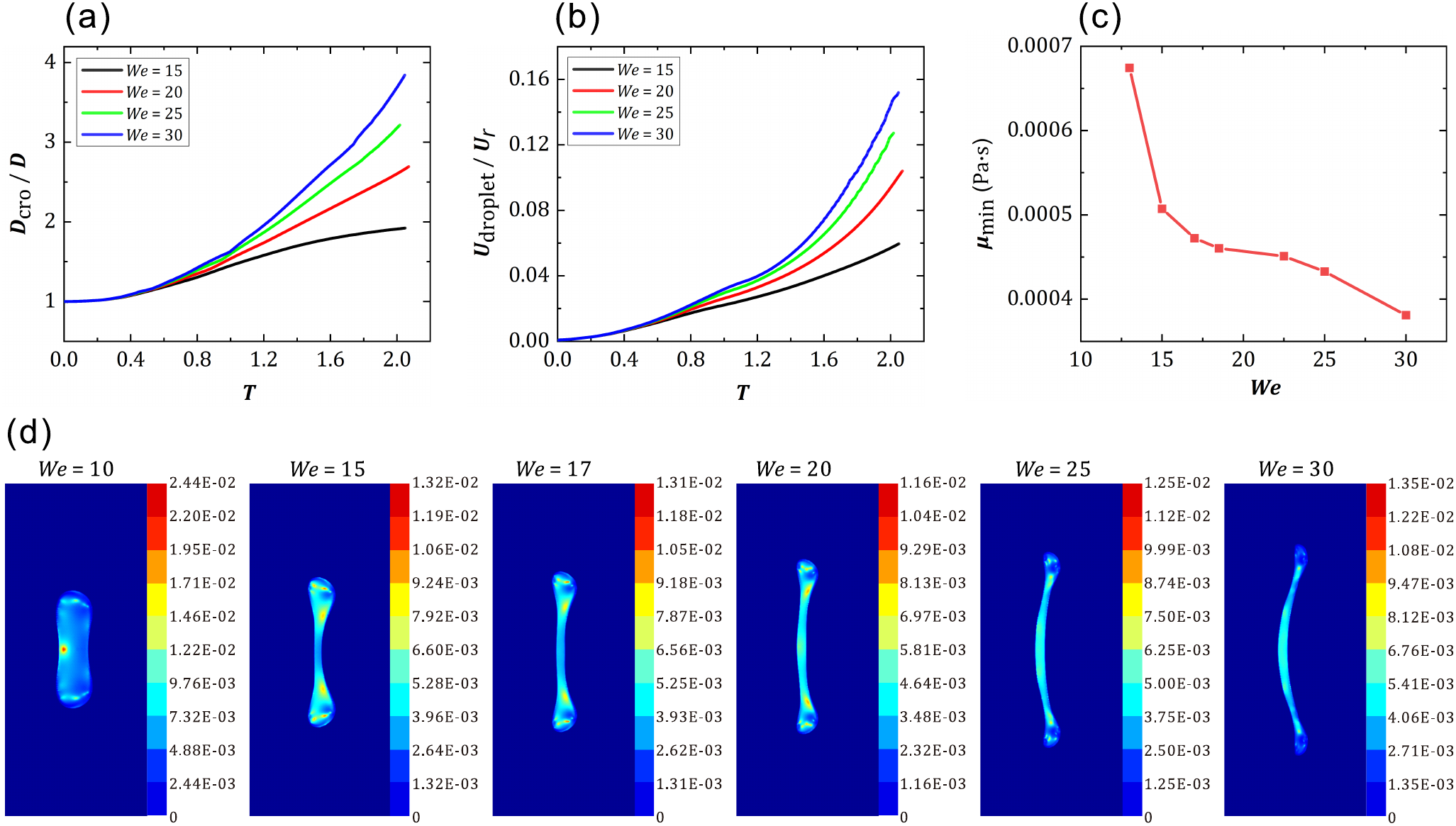}}
  \caption{Evolution of (a) the cross-stream diameter ${{D}_{\rm{cro}}}/D$ and (b) the centroid velocity ${{U}_{\rm{droplet}}}/{{U}_{{r}}}$ of droplets. (c) Minimum effective viscosity ${{\mu }_{\min }}$ and (d) viscosity distribution inside the shear-thinning droplet for various $\We$ values at $T$ = 1.46. Here, $Oh$ = 0.010746 and $n$ = 0.5. }
\label{fig:fig13}
\end{figure*}

To further see the effect of the Weber number on the droplet deformation and breakup quantitatively, we vary the droplet Weber number and analyze the cross-stream diameter and the centroid velocity of the droplet, as shown in Fig.\ \ref{fig:fig13}. The change rates of ${{D}_{\rm{cro}}}/D$ and ${{U}_{\rm{droplet}}}/{{U}_{{r}}}$ increase significantly as $\We$ increases, as shown in Figs.\ \ref{fig:fig13}(a) and \ref{fig:fig13}(b). This is because of the stronger effect of the inertia, the differences between high-viscosity and low-viscosity regions are significant and ${{\mu }_{\min }}$ decreases with the increase of $\We$, as shown in Fig.\ \ref{fig:fig13}(c), which accelerates the droplet deformation. The distribution of the effective viscosity is shown in Fig.\ \ref{fig:fig13}(d). We can see that, as the Weber number increases, the low-viscosity region gradually transits from the center region to the rim region of the droplet. As the low-viscosity region moves, the deformation of the droplet is also affected.

\subsection{Effect of Ohnesorge number}\label{sec:sec034}
\begin{figure}
  \centerline{\includegraphics[width=0.8\columnwidth]{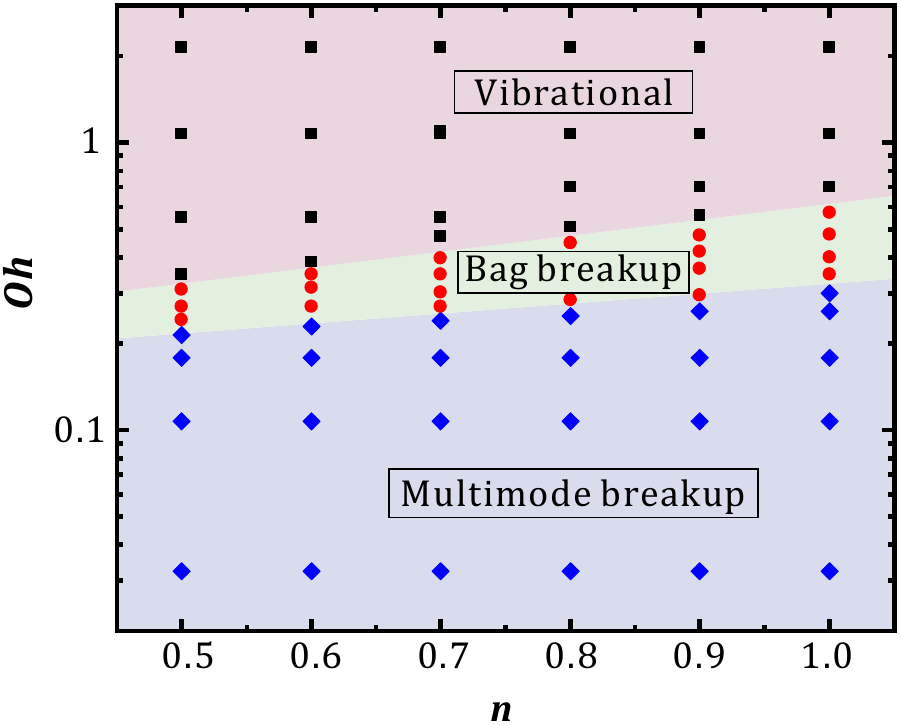}}
  \caption{Regime map for shear-thinning droplets in the $Oh$--$n$ space. Here, $\We$ = 30.}
\label{fig:fig14}
\end{figure}

The effects of the Ohnesorge number $Oh$ on the deformation process and breakup mode of shear-thinning droplets are analyzed in this section. To consider the effect of $Oh$, we fix other parameters ($\We$ and $n$) and change the consistency index $K$ to explore its influence. Then, we vary $Oh$ and $n$ to produce a regime map in the $Oh$--$n$ space, as shown in Fig.\ \ref{fig:fig14}. Three modes occur as $n$ and $Oh$ vary, namely multimode breakup, bag breakup, and vibrational mode. When $Oh$ is small and $n$ is large, the breakup is a multimode breakup, which is mainly because the effective viscosity is small and the droplet deforms easily. As $Oh$ increases, the low-viscosity region transfers from the rim to the center of the droplet, and the breakup is in the bag breakup mode. When $Oh$ is large and $n$ is small, the effective viscosity is very large. Hence, the droplet only vibrates instead of breakup. From Fig.\ \ref{fig:fig14}, we can also see that as $n$ increases, the transitional $Oh$ for the change in the breakup mode increases. This is mainly because when $n$ is large, the value of $K$ needs to be increased to reach the same breakup mode, then the $Oh$ number increases accordingly.

\begin{figure*}
  \centerline{\includegraphics[width=1.8\columnwidth]{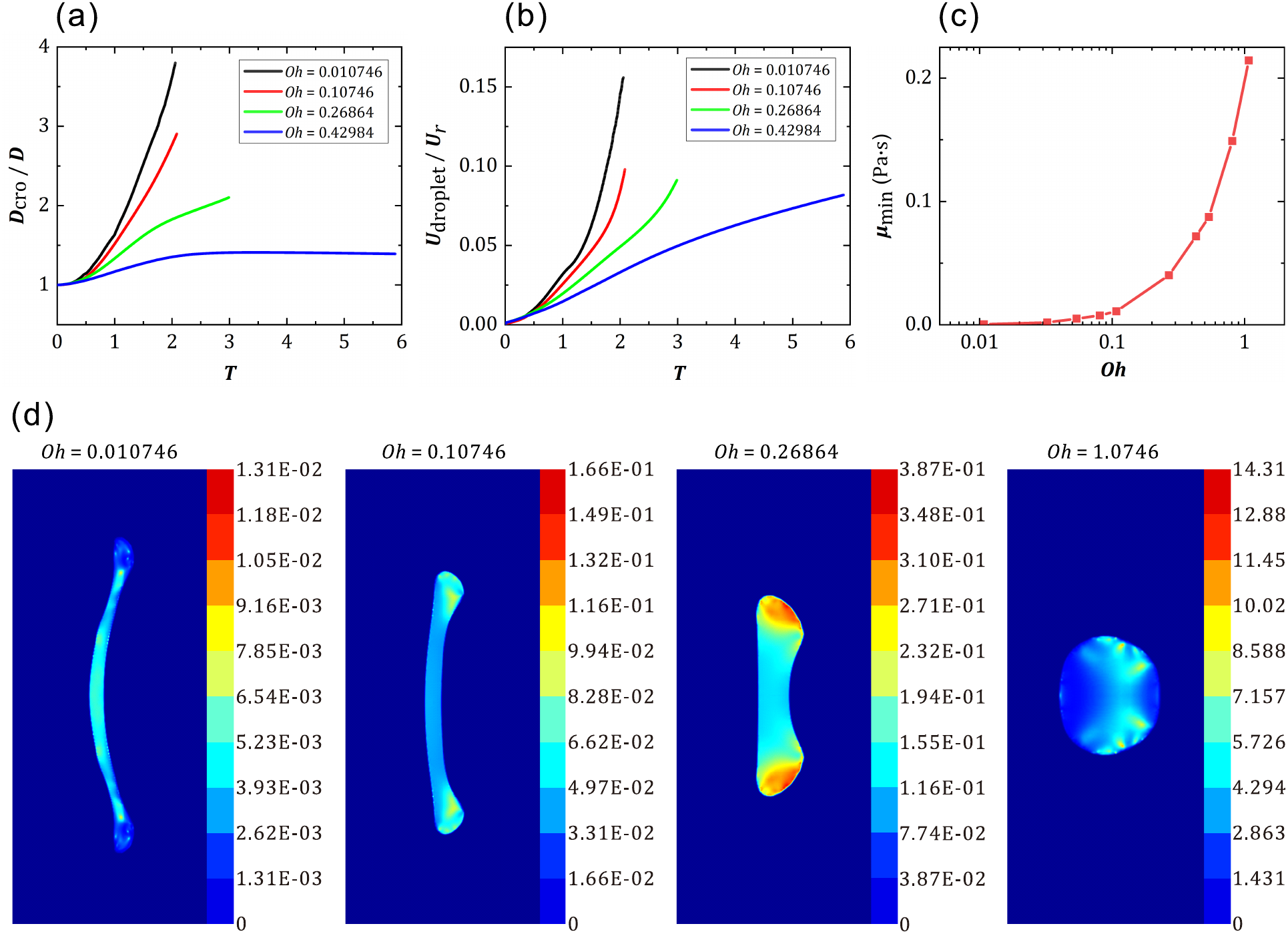}}
  \caption{Evolution of (a) the cross-stream diameter ${{D}_{\rm{cro}}}/D$ and (b) the centroid velocity ${{U}_{\rm{droplet}}}/{{U}_{{r}}}$ of droplets. (c) Minimum effective viscosity ${{\mu }_{\min }}$ and (d) viscosity distribution inside the droplets for various $Oh$ values at $T$ = 1.46. Here, $\We$ = 30 and $n$ = 0.5.}
\label{fig:fig15}
\end{figure*}

To study the effect of the Ohnesorge number on droplet deformation and breakup quantitatively, we vary the Ohnesorge number and analyze the cross-stream diameter and the centroid velocity of the droplet. As $Oh$ increases, the change rates of ${{D}_{\rm{cro}}}/D$ and ${{U}_{\rm{droplet}}}/{{U}_{{r}}}$ of droplets decrease significantly, as shown in Figs.\ \ref{fig:fig15}(a) and \ref{fig:fig15}(b). This is because of the higher viscous stress within the droplet at large $Oh$, which inhibits the droplet deformation remarkably. To confirm this, the minimum effective viscosity ${{\mu }_{\min }}$ for different $Oh$ at $T$ = 1.97 is given in Fig.\ \ref{fig:fig15}(c). When $Oh$ is small and as it increases, the aerodynamic force dominates the deformation of the droplet and the minimum effective viscosity ${{\mu }_{\min }}$ has a small increase. With the further increase of $Oh$, ${{\mu }_{\min }}$ significantly increases as shown in Fig.\ \ref{fig:fig15}(c). In addition, the distribution of the effective viscosity within the shear-thinning droplets is plotted in Fig.\ \ref{fig:fig15}(d). With the increases of $Oh$, the effective viscosity of the droplet increases, and the low-viscosity region inside the droplet transits from the rim of the droplet to the center. Hence, a high-viscosity region appears on the leeward side of the droplet and prohibits the deformation of the droplet. Therefore, the deformation and breakup mode of droplets are also affected.

\section{Conclusion}\label{sec:sec04}
In this study, we consider the secondary breakup of shear-thinning droplets by numerical simulation. To compare between shear-thinning droplets and Newtonian droplets, a new definition of the Ohnesorge number is proposed by considering the characteristic shear rate in the droplet induced by the airflow. The cross-stream diameter and the centroid velocity of the droplet are used to quantify the deformation process of the droplet. The results show that the shear-thinning property can affect the droplet deformation both in the bag breakup and multimode breakup regimes by influencing the local effective viscosity within the droplet. During the bag breakup, as the power-law index $n$ decreases there has small local effective viscosity near the center of the droplets, which accelerates the droplet deformation near the center remarkably. During the multimode breakup, as $n$ decreases, the low viscosity region transits from the center to the rim of droplets and the radial stretching deformation of the droplet is significantly enhanced. Under the same effective viscosity, as the Weber number increases, the minimum effective viscosity decreases, and the breakup transits from the vibrational to the bag breakup, and then to the multimode breakup. As the Ohnesorge number increases, the effective viscosity increases and the deformation degree of the droplets is significantly inhibited. This study mainly considers the deformation and breakup of shear-thinning droplets. Other types of non-Newtonian fluids can affect the process, such as shear-thickening fluids, Bingham fluids, and viscoelastic fluids. The method used in this study can also be extended to consider the non-Newtonian rheology of other fluids, which requires further studies. The results of this study are not only useful for unveiling the influence mechanism of shear-thinning properties in the process of the secondary breakup, but also are helpful to relevant applications such as fuel atomization, spray painting, and powder production.

\section*{Supplementary Material}
See supplementary material for the effect of the initial distance of the droplet from the left boundary.


\section*{Acknowledgements}
This work is supported by the National Natural Science Foundation of China (Grant nos.\ 51676137 and 51921004).

\section*{Data Availability Statement}
The data that support the findings of this study are available from the corresponding author upon reasonable request.

\bibliography{Reference}

\begin{thebibliography}{47}%
\makeatletter
\providecommand \@ifxundefined [1]{%
 \@ifx{#1\undefined}
}%
\providecommand \@ifnum [1]{%
 \ifnum #1\expandafter \@firstoftwo
 \else \expandafter \@secondoftwo
 \fi
}%
\providecommand \@ifx [1]{%
 \ifx #1\expandafter \@firstoftwo
 \else \expandafter \@secondoftwo
 \fi
}%
\providecommand \natexlab [1]{#1}%
\providecommand \enquote  [1]{``#1''}%
\providecommand \bibnamefont  [1]{#1}%
\providecommand \bibfnamefont [1]{#1}%
\providecommand \citenamefont [1]{#1}%
\providecommand \href@noop [0]{\@secondoftwo}%
\providecommand \href [0]{\begingroup \@sanitize@url \@href}%
\providecommand \@href[1]{\@@startlink{#1}\@@href}%
\providecommand \@@href[1]{\endgroup#1\@@endlink}%
\providecommand \@sanitize@url [0]{\catcode `\\12\catcode `\$12\catcode
  `\&12\catcode `\#12\catcode `\^12\catcode `\_12\catcode `\%12\relax}%
\providecommand \@@startlink[1]{}%
\providecommand \@@endlink[0]{}%
\providecommand \url  [0]{\begingroup\@sanitize@url \@url }%
\providecommand \@url [1]{\endgroup\@href {#1}{\urlprefix }}%
\providecommand \urlprefix  [0]{URL }%
\providecommand \Eprint [0]{\href }%
\providecommand \doibase [0]{http://dx.doi.org/}%
\providecommand \selectlanguage [0]{\@gobble}%
\providecommand \bibinfo  [0]{\@secondoftwo}%
\providecommand \bibfield  [0]{\@secondoftwo}%
\providecommand \translation [1]{[#1]}%
\providecommand \BibitemOpen [0]{}%
\providecommand \bibitemStop [0]{}%
\providecommand \bibitemNoStop [0]{.\EOS\space}%
\providecommand \EOS [0]{\spacefactor3000\relax}%
\providecommand \BibitemShut  [1]{\csname bibitem#1\endcsname}%
\let\auto@bib@innerbib\@empty
\bibitem [{\citenamefont {Natan}\ and\ \citenamefont
  {Rahimi}(2002)}]{Natan2002GelPropellants}%
  \BibitemOpen
  \bibfield  {author} {\bibinfo {author} {\bibfnamefont {B.}~\bibnamefont
  {Natan}}\ and\ \bibinfo {author} {\bibfnamefont {S.}~\bibnamefont {Rahimi}},\
  }\bibfield  {title} {\enquote {\bibinfo {title} {The status of gel
  propellants in year 2000},}\ }\href {\doibase
  10.1615/IntJEnergeticMaterialsChemProp.v5.i1-6.200} {\bibfield  {journal}
  {\bibinfo  {journal} {International Journal of Energetic Materials and
  Chemical Propulsion}\ }\textbf {\bibinfo {volume} {5}},\ \bibinfo {pages}
  {172--194} (\bibinfo {year} {2002})}\BibitemShut {NoStop}%
\bibitem [{\citenamefont {Padwal}, \citenamefont {Natan},\ and\ \citenamefont
  {Mishra}(2021)}]{Manisha2021Gel}%
  \BibitemOpen
  \bibfield  {author} {\bibinfo {author} {\bibfnamefont {M.~B.}\ \bibnamefont
  {Padwal}}, \bibinfo {author} {\bibfnamefont {B.}~\bibnamefont {Natan}}, \
  and\ \bibinfo {author} {\bibfnamefont {D.~P.}\ \bibnamefont {Mishra}},\
  }\bibfield  {title} {\enquote {\bibinfo {title} {Gel propellants},}\ }\href
  {\doibase 10.1016/j.pecs.2020.100885} {\bibfield  {journal} {\bibinfo
  {journal} {Prog. Energ. Combust.}\ }\textbf {\bibinfo {volume} {83}},\
  \bibinfo {pages} {100885} (\bibinfo {year} {2021})}\BibitemShut {NoStop}%
\bibitem [{\citenamefont {Cao}\ \emph {et~al.}(2007)\citenamefont {Cao},
  \citenamefont {Sun}, \citenamefont {Li}, \citenamefont {Liu},\ and\
  \citenamefont {Yu}}]{Cao2007Identified}%
  \BibitemOpen
  \bibfield  {author} {\bibinfo {author} {\bibfnamefont {X.~K.}\ \bibnamefont
  {Cao}}, \bibinfo {author} {\bibfnamefont {Z.~G.}\ \bibnamefont {Sun}},
  \bibinfo {author} {\bibfnamefont {W.~F.}\ \bibnamefont {Li}}, \bibinfo
  {author} {\bibfnamefont {H.~F.}\ \bibnamefont {Liu}}, \ and\ \bibinfo
  {author} {\bibfnamefont {Z.~H.}\ \bibnamefont {Yu}},\ }\bibfield  {title}
  {\enquote {\bibinfo {title} {A new breakup regime of liquid drops identified
  in a continuous and uniform air jet flow},}\ }\href {\doibase
  10.1063/1.2723154} {\bibfield  {journal} {\bibinfo  {journal} {Phys. Fluids}\
  }\textbf {\bibinfo {volume} {19}},\ \bibinfo {pages} {057103} (\bibinfo
  {year} {2007})}\BibitemShut {NoStop}%
\bibitem [{\citenamefont {Dai}\ and\ \citenamefont
  {Faeth}(2001)}]{Dai2001PulsedHolography}%
  \BibitemOpen
  \bibfield  {author} {\bibinfo {author} {\bibfnamefont {Z.}~\bibnamefont
  {Dai}}\ and\ \bibinfo {author} {\bibfnamefont {G.}~\bibnamefont {Faeth}},\
  }\bibfield  {title} {\enquote {\bibinfo {title} {Temporal properties of
  secondary drop breakup in the multimode breakup regime},}\ }\href {\doibase
  doi.org/10.1016/S0301-9322(00)00015-X} {\bibfield  {journal} {\bibinfo
  {journal} {Int. J. Multiphase Flow}\ }\textbf {\bibinfo {volume} {27}},\
  \bibinfo {pages} {217--236} (\bibinfo {year} {2001})}\BibitemShut {NoStop}%
\bibitem [{\citenamefont {Joseph}, \citenamefont {Beavers},\ and\ \citenamefont
  {Funada}(2002)}]{Joseph2002RayleighTaylor}%
  \BibitemOpen
  \bibfield  {author} {\bibinfo {author} {\bibfnamefont {D.~D.}\ \bibnamefont
  {Joseph}}, \bibinfo {author} {\bibfnamefont {G.~S.}\ \bibnamefont {Beavers}},
  \ and\ \bibinfo {author} {\bibfnamefont {T.}~\bibnamefont {Funada}},\
  }\bibfield  {title} {\enquote {\bibinfo {title} {{Rayleigh-Taylor}
  instability of viscoelastic drops at high {Weber} numbers},}\ }\href
  {\doibase 10.1017/s0022112001006802} {\bibfield  {journal} {\bibinfo
  {journal} {J. Fluid. Mech.}\ }\textbf {\bibinfo {volume} {453}},\ \bibinfo
  {pages} {109--132} (\bibinfo {year} {2002})}\BibitemShut {NoStop}%
\bibitem [{\citenamefont {Kulkarni}\ and\ \citenamefont
  {Sojka}(2014)}]{Kulkarni2014BagBreakup}%
  \BibitemOpen
  \bibfield  {author} {\bibinfo {author} {\bibfnamefont {V.}~\bibnamefont
  {Kulkarni}}\ and\ \bibinfo {author} {\bibfnamefont {P.~E.}\ \bibnamefont
  {Sojka}},\ }\bibfield  {title} {\enquote {\bibinfo {title} {Bag breakup of
  low viscosity drops in the presence of a continuous air jet},}\ }\href
  {\doibase 10.1063/1.4887817} {\bibfield  {journal} {\bibinfo  {journal}
  {Phys. Fluids}\ }\textbf {\bibinfo {volume} {26}},\ \bibinfo {pages} {072103}
  (\bibinfo {year} {2014})}\BibitemShut {NoStop}%
\bibitem [{\citenamefont {Opfer}\ \emph {et~al.}(2014)\citenamefont {Opfer},
  \citenamefont {Roisman}, \citenamefont {Venzmer}, \citenamefont
  {Klostermann},\ and\ \citenamefont {Tropea}}]{Opfer2014FilmThickness}%
  \BibitemOpen
  \bibfield  {author} {\bibinfo {author} {\bibfnamefont {L.}~\bibnamefont
  {Opfer}}, \bibinfo {author} {\bibfnamefont {I.~V.}\ \bibnamefont {Roisman}},
  \bibinfo {author} {\bibfnamefont {J.}~\bibnamefont {Venzmer}}, \bibinfo
  {author} {\bibfnamefont {M.}~\bibnamefont {Klostermann}}, \ and\ \bibinfo
  {author} {\bibfnamefont {C.}~\bibnamefont {Tropea}},\ }\bibfield  {title}
  {\enquote {\bibinfo {title} {Droplet-air collision dynamics: evolution of the
  film thickness},}\ }\href {\doibase 10.1103/PhysRevE.89.013023} {\bibfield
  {journal} {\bibinfo  {journal} {Phys. Rev. E.}\ }\textbf {\bibinfo {volume}
  {89}},\ \bibinfo {pages} {013023} (\bibinfo {year} {2014})}\BibitemShut
  {NoStop}%
\bibitem [{\citenamefont {Xu}, \citenamefont {Wang},\ and\ \citenamefont
  {Che}(2020)}]{Xu2020ShearFlow}%
  \BibitemOpen
  \bibfield  {author} {\bibinfo {author} {\bibfnamefont {Z.}~\bibnamefont
  {Xu}}, \bibinfo {author} {\bibfnamefont {T.}~\bibnamefont {Wang}}, \ and\
  \bibinfo {author} {\bibfnamefont {Z.}~\bibnamefont {Che}},\ }\bibfield
  {title} {\enquote {\bibinfo {title} {Droplet deformation and breakup in shear
  flow of air},}\ }\href {\doibase 10.1063/5.0006236} {\bibfield  {journal}
  {\bibinfo  {journal} {Phys. Fluids}\ }\textbf {\bibinfo {volume} {32}},\
  \bibinfo {pages} {052109} (\bibinfo {year} {2020})}\BibitemShut {NoStop}%
\bibitem [{\citenamefont {Zhao}\ \emph {et~al.}(2010)\citenamefont {Zhao},
  \citenamefont {Liu}, \citenamefont {Li},\ and\ \citenamefont
  {Xu}}]{Zhao2010MorphologicalClassification}%
  \BibitemOpen
  \bibfield  {author} {\bibinfo {author} {\bibfnamefont {H.}~\bibnamefont
  {Zhao}}, \bibinfo {author} {\bibfnamefont {H.}~\bibnamefont {Liu}}, \bibinfo
  {author} {\bibfnamefont {W.}~\bibnamefont {Li}}, \ and\ \bibinfo {author}
  {\bibfnamefont {J.}~\bibnamefont {Xu}},\ }\bibfield  {title} {\enquote
  {\bibinfo {title} {Morphological classification of low viscosity drop bag
  breakup in a continuous air jet stream},}\ }\href {\doibase
  10.1063/1.3490408} {\bibfield  {journal} {\bibinfo  {journal} {Phys. Fluids}\
  }\textbf {\bibinfo {volume} {22}},\ \bibinfo {pages} {114103} (\bibinfo
  {year} {2010})}\BibitemShut {NoStop}%
\bibitem [{\citenamefont {Zhao}\ \emph {et~al.}(2013)\citenamefont {Zhao},
  \citenamefont {Liu}, \citenamefont {Xu}, \citenamefont {Li},\ and\
  \citenamefont {Lin}}]{Zhao2013TemporalProperties}%
  \BibitemOpen
  \bibfield  {author} {\bibinfo {author} {\bibfnamefont {H.}~\bibnamefont
  {Zhao}}, \bibinfo {author} {\bibfnamefont {H.}~\bibnamefont {Liu}}, \bibinfo
  {author} {\bibfnamefont {J.}~\bibnamefont {Xu}}, \bibinfo {author}
  {\bibfnamefont {W.}~\bibnamefont {Li}}, \ and\ \bibinfo {author}
  {\bibfnamefont {K.}~\bibnamefont {Lin}},\ }\bibfield  {title} {\enquote
  {\bibinfo {title} {Temporal properties of secondary drop breakup in the
  bag-stamen breakup regime},}\ }\href {\doibase 10.1063/1.4803154} {\bibfield
  {journal} {\bibinfo  {journal} {Phys. Fluids}\ }\textbf {\bibinfo {volume}
  {25}},\ \bibinfo {pages} {054102} (\bibinfo {year} {2013})}\BibitemShut
  {NoStop}%
\bibitem [{\citenamefont {Zhao}\ \emph {et~al.}(2016)\citenamefont {Zhao},
  \citenamefont {Zhang}, \citenamefont {Xu}, \citenamefont {Li},\ and\
  \citenamefont {Liu}}]{Zhang2016Surfactant}%
  \BibitemOpen
  \bibfield  {author} {\bibinfo {author} {\bibfnamefont {H.}~\bibnamefont
  {Zhao}}, \bibinfo {author} {\bibfnamefont {W.}~\bibnamefont {Zhang}},
  \bibinfo {author} {\bibfnamefont {J.}~\bibnamefont {Xu}}, \bibinfo {author}
  {\bibfnamefont {W.}~\bibnamefont {Li}}, \ and\ \bibinfo {author}
  {\bibfnamefont {H.}~\bibnamefont {Liu}},\ }\bibfield  {title} {\enquote
  {\bibinfo {title} {Influence of surfactant on the drop bag breakup in a
  continuous air jet stream},}\ }\href {\doibase 10.1063/1.4947575} {\bibfield
  {journal} {\bibinfo  {journal} {Phys. Fluids}\ }\textbf {\bibinfo {volume}
  {28}},\ \bibinfo {pages} {054102} (\bibinfo {year} {2016})}\BibitemShut
  {NoStop}%
\bibitem [{\citenamefont {Dorschner}\ \emph {et~al.}(2020)\citenamefont
  {Dorschner}, \citenamefont {Biasiori-Poulanges}, \citenamefont {Schmidmayer},
  \citenamefont {El-Rabii},\ and\ \citenamefont
  {Colonius}}]{Dorschner2020TransverseRTinstability}%
  \BibitemOpen
  \bibfield  {author} {\bibinfo {author} {\bibfnamefont {B.}~\bibnamefont
  {Dorschner}}, \bibinfo {author} {\bibfnamefont {L.}~\bibnamefont
  {Biasiori-Poulanges}}, \bibinfo {author} {\bibfnamefont {K.}~\bibnamefont
  {Schmidmayer}}, \bibinfo {author} {\bibfnamefont {H.}~\bibnamefont
  {El-Rabii}}, \ and\ \bibinfo {author} {\bibfnamefont {T.}~\bibnamefont
  {Colonius}},\ }\bibfield  {title} {\enquote {\bibinfo {title} {On the
  formation and recurrent shedding of ligaments in droplet aerobreakup},}\
  }\href {\doibase 10.1017/jfm.2020.699} {\bibfield  {journal} {\bibinfo
  {journal} {J. Fluid. Mech.}\ }\textbf {\bibinfo {volume} {904}},\ \bibinfo
  {pages} {A20} (\bibinfo {year} {2020})}\BibitemShut {NoStop}%
\bibitem [{\citenamefont {Feng}(2010)}]{Feng2010InterfacialFlows}%
  \BibitemOpen
  \bibfield  {author} {\bibinfo {author} {\bibfnamefont {J.~Q.}\ \bibnamefont
  {Feng}},\ }\bibfield  {title} {\enquote {\bibinfo {title} {A deformable
  liquid drop falling through a quiescent gas at terminal velocity},}\ }\href
  {\doibase 10.1017/s0022112010001825} {\bibfield  {journal} {\bibinfo
  {journal} {J. Fluid. Mech.}\ }\textbf {\bibinfo {volume} {658}},\ \bibinfo
  {pages} {438--462} (\bibinfo {year} {2010})}\BibitemShut {NoStop}%
\bibitem [{\citenamefont {Jain}\ \emph {et~al.}(2015)\citenamefont {Jain},
  \citenamefont {Prakash}, \citenamefont {Tomar},\ and\ \citenamefont
  {Ravikrishna}}]{Jain2015HighDensityRatio}%
  \BibitemOpen
  \bibfield  {author} {\bibinfo {author} {\bibfnamefont {M.}~\bibnamefont
  {Jain}}, \bibinfo {author} {\bibfnamefont {R.~S.}\ \bibnamefont {Prakash}},
  \bibinfo {author} {\bibfnamefont {G.}~\bibnamefont {Tomar}}, \ and\ \bibinfo
  {author} {\bibfnamefont {R.~V.}\ \bibnamefont {Ravikrishna}},\ }\bibfield
  {title} {\enquote {\bibinfo {title} {Secondary breakup of a drop at moderate
  {Weber} numbers},}\ }\href {\doibase 10.1098/rspa.2014.0930} {\bibfield
  {journal} {\bibinfo  {journal} {Proc. R. Soc. A.}\ }\textbf {\bibinfo
  {volume} {471}},\ \bibinfo {pages} {20140930} (\bibinfo {year}
  {2015})}\BibitemShut {NoStop}%
\bibitem [{\citenamefont {Jain}\ \emph {et~al.}(2019)\citenamefont {Jain},
  \citenamefont {Tyagi}, \citenamefont {Prakash}, \citenamefont {Ravikrishna},\
  and\ \citenamefont {Tomar}}]{Jain2019ModerateWeber}%
  \BibitemOpen
  \bibfield  {author} {\bibinfo {author} {\bibfnamefont {S.~S.}\ \bibnamefont
  {Jain}}, \bibinfo {author} {\bibfnamefont {N.}~\bibnamefont {Tyagi}},
  \bibinfo {author} {\bibfnamefont {R.~S.}\ \bibnamefont {Prakash}}, \bibinfo
  {author} {\bibfnamefont {R.~V.}\ \bibnamefont {Ravikrishna}}, \ and\ \bibinfo
  {author} {\bibfnamefont {G.}~\bibnamefont {Tomar}},\ }\bibfield  {title}
  {\enquote {\bibinfo {title} {Secondary breakup of drops at moderate {Weber}
  numbers: effect of density ratio and {Reynolds} number},}\ }\href {\doibase
  10.1016/j.ijmultiphaseflow.2019.04.026} {\bibfield  {journal} {\bibinfo
  {journal} {Int. J. Multiphase Flow}\ }\textbf {\bibinfo {volume} {117}},\
  \bibinfo {pages} {25--41} (\bibinfo {year} {2019})}\BibitemShut {NoStop}%
\bibitem [{\citenamefont {Jiao}\ \emph {et~al.}(2019)\citenamefont {Jiao},
  \citenamefont {Jiao}, \citenamefont {Zhang},\ and\ \citenamefont
  {Du}}]{Jiao2019TurbulentFlows}%
  \BibitemOpen
  \bibfield  {author} {\bibinfo {author} {\bibfnamefont {D.}~\bibnamefont
  {Jiao}}, \bibinfo {author} {\bibfnamefont {K.}~\bibnamefont {Jiao}}, \bibinfo
  {author} {\bibfnamefont {F.}~\bibnamefont {Zhang}}, \ and\ \bibinfo {author}
  {\bibfnamefont {Q.}~\bibnamefont {Du}},\ }\bibfield  {title} {\enquote
  {\bibinfo {title} {Direct numerical simulation of droplet deformation in
  turbulent flows with different velocity profiles},}\ }\href {\doibase
  10.1016/j.fuel.2019.03.010} {\bibfield  {journal} {\bibinfo  {journal}
  {Fuel}\ }\textbf {\bibinfo {volume} {247}},\ \bibinfo {pages} {302--314}
  (\bibinfo {year} {2019})}\BibitemShut {NoStop}%
\bibitem [{\citenamefont {Ling}, \citenamefont {Zhong},\ and\ \citenamefont
  {Peng}(2021)}]{Ling2021LateralPulsating}%
  \BibitemOpen
  \bibfield  {author} {\bibinfo {author} {\bibfnamefont {C.}~\bibnamefont
  {Ling}}, \bibinfo {author} {\bibfnamefont {Y.}~\bibnamefont {Zhong}}, \ and\
  \bibinfo {author} {\bibfnamefont {L.}~\bibnamefont {Peng}},\ }\bibfield
  {title} {\enquote {\bibinfo {title} {Three-dimensional numerical research on
  the effects of lateral pulsating airflow on droplet breakup},}\ }\href
  {\doibase 10.1063/5.0035051} {\bibfield  {journal} {\bibinfo  {journal}
  {Phys. Fluids}\ }\textbf {\bibinfo {volume} {33}},\ \bibinfo {pages} {033303}
  (\bibinfo {year} {2021})}\BibitemShut {NoStop}%
\bibitem [{\citenamefont {Marcotte}\ and\ \citenamefont
  {Zaleski}(2019)}]{Marcotte2019Thresholds}%
  \BibitemOpen
  \bibfield  {author} {\bibinfo {author} {\bibfnamefont {F.}~\bibnamefont
  {Marcotte}}\ and\ \bibinfo {author} {\bibfnamefont {S.}~\bibnamefont
  {Zaleski}},\ }\bibfield  {title} {\enquote {\bibinfo {title} {Density
  contrast matters for drop fragmentation thresholds at low {Ohnesorge}
  number},}\ }\href {\doibase 10.1103/PhysRevFluids.4.103604} {\bibfield
  {journal} {\bibinfo  {journal} {Phys. Rev. Fluids}\ }\textbf {\bibinfo
  {volume} {4}},\ \bibinfo {pages} {103604} (\bibinfo {year}
  {2019})}\BibitemShut {NoStop}%
\bibitem [{\citenamefont {Meng}\ and\ \citenamefont
  {Colonius}(2017)}]{Meng2018SheetInstability}%
  \BibitemOpen
  \bibfield  {author} {\bibinfo {author} {\bibfnamefont {J.~C.}\ \bibnamefont
  {Meng}}\ and\ \bibinfo {author} {\bibfnamefont {T.}~\bibnamefont
  {Colonius}},\ }\bibfield  {title} {\enquote {\bibinfo {title} {Numerical
  simulation of the aerobreakup of a water droplet},}\ }\href {\doibase
  10.1017/jfm.2017.804} {\bibfield  {journal} {\bibinfo  {journal} {J. Fluid.
  Mech.}\ }\textbf {\bibinfo {volume} {835}},\ \bibinfo {pages} {1108--1135}
  (\bibinfo {year} {2017})}\BibitemShut {NoStop}%
\bibitem [{\citenamefont {Obenauf}\ and\ \citenamefont
  {Sojka}(2021)}]{Sojka2021MultimodeBreakup}%
  \BibitemOpen
  \bibfield  {author} {\bibinfo {author} {\bibfnamefont {D.~G.}\ \bibnamefont
  {Obenauf}}\ and\ \bibinfo {author} {\bibfnamefont {P.~E.}\ \bibnamefont
  {Sojka}},\ }\bibfield  {title} {\enquote {\bibinfo {title} {Theoretical
  deformation modeling and drop size prediction in the multimode breakup
  regime},}\ }\href {\doibase 10.1063/5.0062040} {\bibfield  {journal}
  {\bibinfo  {journal} {Phys. Fluids}\ }\textbf {\bibinfo {volume} {33}},\
  \bibinfo {pages} {092113} (\bibinfo {year} {2021})}\BibitemShut {NoStop}%
\bibitem [{\citenamefont {Rimbert}\ \emph {et~al.}(2020)\citenamefont
  {Rimbert}, \citenamefont {Castrillon~Escobar}, \citenamefont {Meignen},
  \citenamefont {Hadj-Achour},\ and\ \citenamefont
  {Gradeck}}]{Rimbert2020DropletDeformation}%
  \BibitemOpen
  \bibfield  {author} {\bibinfo {author} {\bibfnamefont {N.}~\bibnamefont
  {Rimbert}}, \bibinfo {author} {\bibfnamefont {S.}~\bibnamefont
  {Castrillon~Escobar}}, \bibinfo {author} {\bibfnamefont {R.}~\bibnamefont
  {Meignen}}, \bibinfo {author} {\bibfnamefont {M.}~\bibnamefont
  {Hadj-Achour}}, \ and\ \bibinfo {author} {\bibfnamefont {M.}~\bibnamefont
  {Gradeck}},\ }\bibfield  {title} {\enquote {\bibinfo {title} {Spheroidal
  droplet deformation, oscillation and breakup in uniform outer flow},}\ }\href
  {\doibase 10.1017/jfm.2020.675} {\bibfield  {journal} {\bibinfo  {journal}
  {J. Fluid. Mech.}\ }\textbf {\bibinfo {volume} {904}},\ \bibinfo {pages}
  {A15} (\bibinfo {year} {2020})}\BibitemShut {NoStop}%
\bibitem [{\citenamefont {Stefanitsis}\ \emph {et~al.}(2019)\citenamefont
  {Stefanitsis}, \citenamefont {Malgarinos}, \citenamefont {Strotos},
  \citenamefont {Nikolopoulos}, \citenamefont {Kakaras},\ and\ \citenamefont
  {Gavaises}}]{Stefanitsis2019DropletsInTandem}%
  \BibitemOpen
  \bibfield  {author} {\bibinfo {author} {\bibfnamefont {D.}~\bibnamefont
  {Stefanitsis}}, \bibinfo {author} {\bibfnamefont {I.}~\bibnamefont
  {Malgarinos}}, \bibinfo {author} {\bibfnamefont {G.}~\bibnamefont {Strotos}},
  \bibinfo {author} {\bibfnamefont {N.}~\bibnamefont {Nikolopoulos}}, \bibinfo
  {author} {\bibfnamefont {E.}~\bibnamefont {Kakaras}}, \ and\ \bibinfo
  {author} {\bibfnamefont {M.}~\bibnamefont {Gavaises}},\ }\bibfield  {title}
  {\enquote {\bibinfo {title} {Numerical investigation of the aerodynamic
  breakup of droplets in tandem},}\ }\href {\doibase
  10.1016/j.ijmultiphaseflow.2018.10.015} {\bibfield  {journal} {\bibinfo
  {journal} {Int. J. Multiphase Flow}\ }\textbf {\bibinfo {volume} {113}},\
  \bibinfo {pages} {289--303} (\bibinfo {year} {2019})}\BibitemShut {NoStop}%
\bibitem [{\citenamefont {Yang}\ \emph {et~al.}(2017)\citenamefont {Yang},
  \citenamefont {Jia}, \citenamefont {Che}, \citenamefont {Sun},\ and\
  \citenamefont {Wang}}]{Yang2017Transitions}%
  \BibitemOpen
  \bibfield  {author} {\bibinfo {author} {\bibfnamefont {W.}~\bibnamefont
  {Yang}}, \bibinfo {author} {\bibfnamefont {M.}~\bibnamefont {Jia}}, \bibinfo
  {author} {\bibfnamefont {Z.}~\bibnamefont {Che}}, \bibinfo {author}
  {\bibfnamefont {K.}~\bibnamefont {Sun}}, \ and\ \bibinfo {author}
  {\bibfnamefont {T.}~\bibnamefont {Wang}},\ }\bibfield  {title} {\enquote
  {\bibinfo {title} {Transitions of deformation to bag breakup and bag to
  bag-stamen breakup for droplets subjected to a continuous gas flow},}\ }\href
  {\doibase 10.1016/j.ijheatmasstransfer.2017.04.012} {\bibfield  {journal}
  {\bibinfo  {journal} {Int. J. Heat Mass Tran.}\ }\textbf {\bibinfo {volume}
  {111}},\ \bibinfo {pages} {884--894} (\bibinfo {year} {2017})}\BibitemShut
  {NoStop}%
\bibitem [{\citenamefont {Yang}\ \emph {et~al.}(2016)\citenamefont {Yang},
  \citenamefont {Jia}, \citenamefont {Sun},\ and\ \citenamefont
  {Wang}}]{Yang2016HighlyUnstable}%
  \BibitemOpen
  \bibfield  {author} {\bibinfo {author} {\bibfnamefont {W.}~\bibnamefont
  {Yang}}, \bibinfo {author} {\bibfnamefont {M.}~\bibnamefont {Jia}}, \bibinfo
  {author} {\bibfnamefont {K.}~\bibnamefont {Sun}}, \ and\ \bibinfo {author}
  {\bibfnamefont {T.}~\bibnamefont {Wang}},\ }\bibfield  {title} {\enquote
  {\bibinfo {title} {Influence of density ratio on the secondary atomization of
  liquid droplets under highly unstable conditions},}\ }\href {\doibase
  10.1016/j.fuel.2016.01.078} {\bibfield  {journal} {\bibinfo  {journal}
  {Fuel}\ }\textbf {\bibinfo {volume} {174}},\ \bibinfo {pages} {25--35}
  (\bibinfo {year} {2016})}\BibitemShut {NoStop}%
\bibitem [{\citenamefont {Zhu}\ \emph {et~al.}(2021)\citenamefont {Zhu},
  \citenamefont {Zhao}, \citenamefont {Jia}, \citenamefont {Chen},\ and\
  \citenamefont {Zheng}}]{Zhu2021AirflowPressure}%
  \BibitemOpen
  \bibfield  {author} {\bibinfo {author} {\bibfnamefont {W.}~\bibnamefont
  {Zhu}}, \bibinfo {author} {\bibfnamefont {N.}~\bibnamefont {Zhao}}, \bibinfo
  {author} {\bibfnamefont {X.}~\bibnamefont {Jia}}, \bibinfo {author}
  {\bibfnamefont {X.}~\bibnamefont {Chen}}, \ and\ \bibinfo {author}
  {\bibfnamefont {H.}~\bibnamefont {Zheng}},\ }\bibfield  {title} {\enquote
  {\bibinfo {title} {Effect of airflow pressure on the droplet breakup in the
  shear breakup regime},}\ }\href {\doibase 10.1063/5.0049558} {\bibfield
  {journal} {\bibinfo  {journal} {Phys. Fluids}\ }\textbf {\bibinfo {volume}
  {33}},\ \bibinfo {pages} {053309} (\bibinfo {year} {2021})}\BibitemShut
  {NoStop}%
\bibitem [{\citenamefont {Gao}\ \emph {et~al.}(2014)\citenamefont {Gao},
  \citenamefont {Rodrigues}, \citenamefont {Sojka},\ and\ \citenamefont
  {Chen}}]{Gao2014Inline}%
  \BibitemOpen
  \bibfield  {author} {\bibinfo {author} {\bibfnamefont {J.}~\bibnamefont
  {Gao}}, \bibinfo {author} {\bibfnamefont {N.~S.}\ \bibnamefont {Rodrigues}},
  \bibinfo {author} {\bibfnamefont {P.~E.}\ \bibnamefont {Sojka}}, \ and\
  \bibinfo {author} {\bibfnamefont {J.}~\bibnamefont {Chen}},\ }\bibfield
  {title} {\enquote {\bibinfo {title} {Measurement of aerodynamic breakup of
  {non-Newtonian} drops by digital in-line holography},}\ }in\ \href {\doibase
  doi.org/10.1115/FEDSM2014-22039} {\emph {\bibinfo {booktitle} {Fluids
  Engineering Division Summer Meeting}}}\ (\bibinfo {year} {2014})\ p.\
  \bibinfo {pages} {V002T11A009}\BibitemShut {NoStop}%
\bibitem [{\citenamefont {Qian}\ \emph {et~al.}(2021)\citenamefont {Qian},
  \citenamefont {Zhong}, \citenamefont {Zhu},\ and\ \citenamefont
  {Lin}}]{Qian2021CarboxymethylCelluloseDroplets}%
  \BibitemOpen
  \bibfield  {author} {\bibinfo {author} {\bibfnamefont {L.}~\bibnamefont
  {Qian}}, \bibinfo {author} {\bibfnamefont {X.}~\bibnamefont {Zhong}},
  \bibinfo {author} {\bibfnamefont {C.}~\bibnamefont {Zhu}}, \ and\ \bibinfo
  {author} {\bibfnamefont {J.}~\bibnamefont {Lin}},\ }\bibfield  {title}
  {\enquote {\bibinfo {title} {An experimental investigation on the secondary
  breakup of carboxymethyl cellulose droplets},}\ }\href {\doibase
  10.1016/j.ijmultiphaseflow.2020.103526} {\bibfield  {journal} {\bibinfo
  {journal} {Int. J. Multiphase Flow}\ }\textbf {\bibinfo {volume} {136}},\
  \bibinfo {pages} {103526} (\bibinfo {year} {2021})}\BibitemShut {NoStop}%
\bibitem [{\citenamefont {Snyder}, \citenamefont {Arockiam},\ and\
  \citenamefont {Sojka}(2010)}]{Snyder2010ElasticNonNewtonian}%
  \BibitemOpen
  \bibfield  {author} {\bibinfo {author} {\bibfnamefont {S.}~\bibnamefont
  {Snyder}}, \bibinfo {author} {\bibfnamefont {N.}~\bibnamefont {Arockiam}}, \
  and\ \bibinfo {author} {\bibfnamefont {P.}~\bibnamefont {Sojka}},\ }\bibfield
   {title} {\enquote {\bibinfo {title} {Secondary atomization of elastic
  {non-Newtonian} liquid drops},}\ }in\ \href@noop {} {\emph {\bibinfo
  {booktitle} {46th AIAA/ASME/SAE/ASEE Joint Propulsion Conference \&
  Exhibit}}}\ (\bibinfo {year} {2010})\ p.\ \bibinfo {pages} {6822}\BibitemShut
  {NoStop}%
\bibitem [{\citenamefont {Wang}\ \emph {et~al.}(2021)\citenamefont {Wang},
  \citenamefont {Zhao}, \citenamefont {Li}, \citenamefont {Xu},\ and\
  \citenamefont {Liu}}]{Wang2021ShearThickening}%
  \BibitemOpen
  \bibfield  {author} {\bibinfo {author} {\bibfnamefont {Z.~Y.}\ \bibnamefont
  {Wang}}, \bibinfo {author} {\bibfnamefont {H.}~\bibnamefont {Zhao}}, \bibinfo
  {author} {\bibfnamefont {W.~F.}\ \bibnamefont {Li}}, \bibinfo {author}
  {\bibfnamefont {J.~L.}\ \bibnamefont {Xu}}, \ and\ \bibinfo {author}
  {\bibfnamefont {H.~F.}\ \bibnamefont {Liu}},\ }\bibfield  {title} {\enquote
  {\bibinfo {title} {Secondary breakup of shear thickening suspension drop},}\
  }\href {\doibase 10.1063/5.0062345} {\bibfield  {journal} {\bibinfo
  {journal} {Phys. Fluids}\ }\textbf {\bibinfo {volume} {33}},\ \bibinfo
  {pages} {093103} (\bibinfo {year} {2021})}\BibitemShut {NoStop}%
\bibitem [{\citenamefont {Zhao}\ \emph {et~al.}(2011)\citenamefont {Zhao},
  \citenamefont {Liu}, \citenamefont {Xu},\ and\ \citenamefont
  {Li}}]{Zhao2011CoalWaterSlurry}%
  \BibitemOpen
  \bibfield  {author} {\bibinfo {author} {\bibfnamefont {H.}~\bibnamefont
  {Zhao}}, \bibinfo {author} {\bibfnamefont {H.}~\bibnamefont {Liu}}, \bibinfo
  {author} {\bibfnamefont {J.}~\bibnamefont {Xu}}, \ and\ \bibinfo {author}
  {\bibfnamefont {W.}~\bibnamefont {Li}},\ }\bibfield  {title} {\enquote
  {\bibinfo {title} {Secondary breakup of coal water slurry drops},}\ }\href
  {\doibase 10.1063/1.3659495} {\bibfield  {journal} {\bibinfo  {journal}
  {Phys. Fluids}\ }\textbf {\bibinfo {volume} {23}},\ \bibinfo {pages} {113101}
  (\bibinfo {year} {2011})}\BibitemShut {NoStop}%
\bibitem [{\citenamefont {Cao}\ \emph {et~al.}(2022)\citenamefont {Cao},
  \citenamefont {Liao}, \citenamefont {Natan}, \citenamefont {Feng},\ and\
  \citenamefont {Wu}}]{Cao2022KeroseneGel}%
  \BibitemOpen
  \bibfield  {author} {\bibinfo {author} {\bibfnamefont {Q.}~\bibnamefont
  {Cao}}, \bibinfo {author} {\bibfnamefont {W.}~\bibnamefont {Liao}}, \bibinfo
  {author} {\bibfnamefont {B.}~\bibnamefont {Natan}}, \bibinfo {author}
  {\bibfnamefont {F.}~\bibnamefont {Feng}}, \ and\ \bibinfo {author}
  {\bibfnamefont {W.}~\bibnamefont {Wu}},\ }\bibfield  {title} {\enquote
  {\bibinfo {title} {Secondary atomization of {non-Newtonian} kerosene gel at
  low {Weber} numbers: A numerical study},}\ }\href {\doibase
  10.1016/j.ast.2021.107280} {\bibfield  {journal} {\bibinfo  {journal}
  {Aerosp. Sci. Technol.}\ }\textbf {\bibinfo {volume} {120}},\ \bibinfo
  {pages} {107280} (\bibinfo {year} {2022})}\BibitemShut {NoStop}%
\bibitem [{\citenamefont {Chu}\ \emph {et~al.}(2020)\citenamefont {Chu},
  \citenamefont {Qian}, \citenamefont {Zhong}, \citenamefont {Zhu},\ and\
  \citenamefont {Chen}}]{Chu2020PolymerSolution}%
  \BibitemOpen
  \bibfield  {author} {\bibinfo {author} {\bibfnamefont {G.}~\bibnamefont
  {Chu}}, \bibinfo {author} {\bibfnamefont {L.}~\bibnamefont {Qian}}, \bibinfo
  {author} {\bibfnamefont {X.}~\bibnamefont {Zhong}}, \bibinfo {author}
  {\bibfnamefont {C.}~\bibnamefont {Zhu}}, \ and\ \bibinfo {author}
  {\bibfnamefont {Z.}~\bibnamefont {Chen}},\ }\bibfield  {title} {\enquote
  {\bibinfo {title} {A numerical investigation on droplet bag breakup behavior
  of polymer solution},}\ }\href {\doibase 10.3390/polym12102172} {\bibfield
  {journal} {\bibinfo  {journal} {Polymers}\ }\textbf {\bibinfo {volume}
  {12}},\ \bibinfo {pages} {2172} (\bibinfo {year} {2020})}\BibitemShut
  {NoStop}%
\bibitem [{\citenamefont {Fu}\ \emph {et~al.}(2019)\citenamefont {Fu},
  \citenamefont {Hou}, \citenamefont {Ren}, \citenamefont {Zhang},
  \citenamefont {Mao},\ and\ \citenamefont {Yu}}]{Fu2019ImpactModel}%
  \BibitemOpen
  \bibfield  {author} {\bibinfo {author} {\bibfnamefont {P.}~\bibnamefont
  {Fu}}, \bibinfo {author} {\bibfnamefont {L.}~\bibnamefont {Hou}}, \bibinfo
  {author} {\bibfnamefont {Z.}~\bibnamefont {Ren}}, \bibinfo {author}
  {\bibfnamefont {Z.}~\bibnamefont {Zhang}}, \bibinfo {author} {\bibfnamefont
  {X.}~\bibnamefont {Mao}}, \ and\ \bibinfo {author} {\bibfnamefont
  {Y.}~\bibnamefont {Yu}},\ }\bibfield  {title} {\enquote {\bibinfo {title} {A
  droplet/wall impact model and simulation of a bipropellant rocket engine},}\
  }\href {\doibase 10.1016/j.ast.2019.03.018} {\bibfield  {journal} {\bibinfo
  {journal} {Aerosp. Sci. Technol.}\ }\textbf {\bibinfo {volume} {88}},\
  \bibinfo {pages} {32--39} (\bibinfo {year} {2019})}\BibitemShut {NoStop}%
\bibitem [{\citenamefont {Kant}\ and\ \citenamefont
  {Banerjee}(2022)}]{Kant2022BreakupMechanism}%
  \BibitemOpen
  \bibfield  {author} {\bibinfo {author} {\bibfnamefont {K.}~\bibnamefont
  {Kant}}\ and\ \bibinfo {author} {\bibfnamefont {R.}~\bibnamefont
  {Banerjee}},\ }\bibfield  {title} {\enquote {\bibinfo {title} {Study of the
  secondary droplet breakup mechanism and regime map of {Newtonian} and power
  law fluids at high liquid-gas density ratio},}\ }\href {\doibase
  10.1063/5.0088144} {\bibfield  {journal} {\bibinfo  {journal} {Phys. Fluids}\
  }\textbf {\bibinfo {volume} {34}},\ \bibinfo {pages} {043108} (\bibinfo
  {year} {2022})}\BibitemShut {NoStop}%
\bibitem [{\citenamefont {Markovich}\ \emph {et~al.}(2019)\citenamefont
  {Markovich}, \citenamefont {Shebeleva}, \citenamefont {Minakov},
  \citenamefont {Lobasov}, \citenamefont {Shebelev}, \citenamefont {Kuibin},\
  and\ \citenamefont {Vorobyev}}]{Markovich2019Destruction}%
  \BibitemOpen
  \bibfield  {author} {\bibinfo {author} {\bibfnamefont {D.~M.}\ \bibnamefont
  {Markovich}}, \bibinfo {author} {\bibfnamefont {A.}~\bibnamefont
  {Shebeleva}}, \bibinfo {author} {\bibfnamefont {A.}~\bibnamefont {Minakov}},
  \bibinfo {author} {\bibfnamefont {A.}~\bibnamefont {Lobasov}}, \bibinfo
  {author} {\bibfnamefont {A.}~\bibnamefont {Shebelev}}, \bibinfo {author}
  {\bibfnamefont {P.~A.}\ \bibnamefont {Kuibin}}, \ and\ \bibinfo {author}
  {\bibfnamefont {M.~A.}\ \bibnamefont {Vorobyev}},\ }\bibfield  {title}
  {\enquote {\bibinfo {title} {Numerical modelling of destruction of a drop of
  non-{N}ewtonian fluid in a gas flow},}\ }\href {\doibase
  10.1051/epjconf/201919600042} {\bibfield  {journal} {\bibinfo  {journal} {EPJ
  Web. Conf.}\ }\textbf {\bibinfo {volume} {196}},\ \bibinfo {pages} {00042}
  (\bibinfo {year} {2019})}\BibitemShut {NoStop}%
\bibitem [{\citenamefont {Minakov}\ \emph {et~al.}(2019)\citenamefont
  {Minakov}, \citenamefont {Shebeleva}, \citenamefont {Strizhak}, \citenamefont
  {Chernetskiy},\ and\ \citenamefont {Volkov}}]{Minakov2019Petrochemicals}%
  \BibitemOpen
  \bibfield  {author} {\bibinfo {author} {\bibfnamefont {A.~V.}\ \bibnamefont
  {Minakov}}, \bibinfo {author} {\bibfnamefont {A.~A.}\ \bibnamefont
  {Shebeleva}}, \bibinfo {author} {\bibfnamefont {P.~A.}\ \bibnamefont
  {Strizhak}}, \bibinfo {author} {\bibfnamefont {M.~Y.}\ \bibnamefont
  {Chernetskiy}}, \ and\ \bibinfo {author} {\bibfnamefont {R.~S.}\ \bibnamefont
  {Volkov}},\ }\bibfield  {title} {\enquote {\bibinfo {title} {Study of the
  {Weber} number impact on secondary breakup of droplets of coal water slurries
  containing petrochemicals},}\ }\href {\doibase 10.1016/j.fuel.2019.06.014}
  {\bibfield  {journal} {\bibinfo  {journal} {Fuel}\ }\textbf {\bibinfo
  {volume} {254}},\ \bibinfo {pages} {115606} (\bibinfo {year}
  {2019})}\BibitemShut {NoStop}%
\bibitem [{\citenamefont {Tavangar}, \citenamefont {Hashemabadi},\ and\
  \citenamefont {Saberimoghadam}(2015)}]{Tavangar2015CoalWaterSlurry}%
  \BibitemOpen
  \bibfield  {author} {\bibinfo {author} {\bibfnamefont {S.}~\bibnamefont
  {Tavangar}}, \bibinfo {author} {\bibfnamefont {S.~H.}\ \bibnamefont
  {Hashemabadi}}, \ and\ \bibinfo {author} {\bibfnamefont {A.}~\bibnamefont
  {Saberimoghadam}},\ }\bibfield  {title} {\enquote {\bibinfo {title} {{CFD}
  simulation for secondary breakup of coal-water slurry drops using
  {OpenFOAM}},}\ }\href {\doibase 10.1016/j.fuproc.2014.12.037} {\bibfield
  {journal} {\bibinfo  {journal} {Fuel Process Technol.}\ }\textbf {\bibinfo
  {volume} {132}},\ \bibinfo {pages} {153--163} (\bibinfo {year}
  {2015})}\BibitemShut {NoStop}%
\bibitem [{\citenamefont {Verhulst}\ \emph {et~al.}(2009)\citenamefont
  {Verhulst}, \citenamefont {Cardinaels}, \citenamefont {Moldenaers},
  \citenamefont {Renardy},\ and\ \citenamefont
  {Afkhami}}]{Verhulst2009BlendMorphology}%
  \BibitemOpen
  \bibfield  {author} {\bibinfo {author} {\bibfnamefont {K.}~\bibnamefont
  {Verhulst}}, \bibinfo {author} {\bibfnamefont {R.}~\bibnamefont
  {Cardinaels}}, \bibinfo {author} {\bibfnamefont {P.}~\bibnamefont
  {Moldenaers}}, \bibinfo {author} {\bibfnamefont {Y.}~\bibnamefont {Renardy}},
  \ and\ \bibinfo {author} {\bibfnamefont {S.}~\bibnamefont {Afkhami}},\
  }\bibfield  {title} {\enquote {\bibinfo {title} {Influence of viscoelasticity
  on drop deformation and orientation in shear flow},}\ }\href {\doibase
  10.1016/j.jnnfm.2008.06.007} {\bibfield  {journal} {\bibinfo  {journal} {J.
  Non-Newton. Fluid Mech.}\ }\textbf {\bibinfo {volume} {156}},\ \bibinfo
  {pages} {29--43} (\bibinfo {year} {2009})}\BibitemShut {NoStop}%
\bibitem [{\citenamefont {Wong}\ \emph {et~al.}(2019)\citenamefont {Wong},
  \citenamefont {Loizou}, \citenamefont {Lau}, \citenamefont {Graham},\ and\
  \citenamefont {Hewakandamby}}]{Wong2019DispersedPhase}%
  \BibitemOpen
  \bibfield  {author} {\bibinfo {author} {\bibfnamefont {V.~L.}\ \bibnamefont
  {Wong}}, \bibinfo {author} {\bibfnamefont {K.}~\bibnamefont {Loizou}},
  \bibinfo {author} {\bibfnamefont {P.~L.}\ \bibnamefont {Lau}}, \bibinfo
  {author} {\bibfnamefont {R.~S.}\ \bibnamefont {Graham}}, \ and\ \bibinfo
  {author} {\bibfnamefont {B.~N.}\ \bibnamefont {Hewakandamby}},\ }\bibfield
  {title} {\enquote {\bibinfo {title} {Characterizing droplet breakup rates of
  shear-thinning dispersed phase in microreactors},}\ }\href {\doibase
  10.1016/j.cherd.2019.02.024} {\bibfield  {journal} {\bibinfo  {journal}
  {Chem. Eng. Res. Des.}\ }\textbf {\bibinfo {volume} {144}},\ \bibinfo {pages}
  {370--385} (\bibinfo {year} {2019})}\BibitemShut {NoStop}%
\bibitem [{\citenamefont {Hirt}\ and\ \citenamefont
  {Nichols}(1981)}]{Hirt1981FreeBoundaries}%
  \BibitemOpen
  \bibfield  {author} {\bibinfo {author} {\bibfnamefont {C.~W.}\ \bibnamefont
  {Hirt}}\ and\ \bibinfo {author} {\bibfnamefont {B.~D.}\ \bibnamefont
  {Nichols}},\ }\bibfield  {title} {\enquote {\bibinfo {title} {Volume of fluid
  ({VOF}) method for the dynamics of free boundaries},}\ }\href {\doibase
  10.1016/0021-9991(81)90145-5} {\bibfield  {journal} {\bibinfo  {journal} {J.
  Comput. Phys.}\ }\textbf {\bibinfo {volume} {39}},\ \bibinfo {pages}
  {201--225} (\bibinfo {year} {1981})}\BibitemShut {NoStop}%
\bibitem [{\citenamefont {Berberovic}(2010)}]{Berberovic2010impact}%
  \BibitemOpen
  \bibfield  {author} {\bibinfo {author} {\bibfnamefont {E.}~\bibnamefont
  {Berberovic}},\ }\emph {\bibinfo {title} {Investigation of free-surface flow
  associated with drop impact: numerical simulations and theoretical
  modeling}},\ \href@noop {} {\bibinfo {type} {Thesis}},\ \bibinfo  {school}
  {Technische Universit\"{a}t} (\bibinfo {year} {2010})\BibitemShut {NoStop}%
\bibitem [{\citenamefont {Arnold}\ \emph {et~al.}(2010)\citenamefont {Arnold},
  \citenamefont {Anderson}, \citenamefont {Santos}, \citenamefont {deRidder},\
  and\ \citenamefont {Campanella}}]{Arnold2010Gels}%
  \BibitemOpen
  \bibfield  {author} {\bibinfo {author} {\bibfnamefont {R.}~\bibnamefont
  {Arnold}}, \bibinfo {author} {\bibfnamefont {W.}~\bibnamefont {Anderson}},
  \bibinfo {author} {\bibfnamefont {P.}~\bibnamefont {Santos}}, \bibinfo
  {author} {\bibfnamefont {M.}~\bibnamefont {deRidder}}, \ and\ \bibinfo
  {author} {\bibfnamefont {O.}~\bibnamefont {Campanella}},\ }\bibfield  {title}
  {\enquote {\bibinfo {title} {Comparison of
  monomethylhydrazine/hydroxypropylcellulose and hydrocarbon/silica gels},}\
  }in\ \href@noop {} {\emph {\bibinfo {booktitle} {48th AIAA Aerospace Sciences
  Meeting Including the New Horizons Forum and Aerospace Exposition}}}\
  (\bibinfo {year} {2010})\ p.\ \bibinfo {pages} {422}\BibitemShut {NoStop}%
\bibitem [{\citenamefont {Popinet}()}]{Popinet2018Basilisk}%
  \BibitemOpen
  \bibfield  {author} {\bibinfo {author} {\bibfnamefont {S.}~\bibnamefont
  {Popinet}},\ }\bibfield  {title} {\enquote {\bibinfo {title} {Basilisk, a
  free-software program for the solution of partial differential equations on
  adaptive cartesian meshes},}\ }\href@noop {} {\bibinfo  {journal}
  {http://basilisk.fr}\ }\BibitemShut {NoStop}%
\bibitem [{\citenamefont {Lagr\'{e}e}, \citenamefont {Staron},\ and\
  \citenamefont {Popinet}(2011)}]{Lagree2011GranularMedia}%
  \BibitemOpen
\bibfield  {journal} {  }\bibfield  {author} {\bibinfo {author} {\bibfnamefont
  {P.-Y.}\ \bibnamefont {Lagr\'{e}e}}, \bibinfo {author} {\bibfnamefont
  {L.}~\bibnamefont {Staron}}, \ and\ \bibinfo {author} {\bibfnamefont
  {S.}~\bibnamefont {Popinet}},\ }\bibfield  {title} {\enquote {\bibinfo
  {title} {The granular column collapse as a continuum: validity of a
  two-dimensional {Navier-Stokes} model with a $\mu$({I})-rheology},}\ }\href
  {\doibase 10.1017/jfm.2011.335} {\bibfield  {journal} {\bibinfo  {journal}
  {J. Fluid. Mech.}\ }\textbf {\bibinfo {volume} {686}},\ \bibinfo {pages}
  {378--408} (\bibinfo {year} {2011})}\BibitemShut {NoStop}%
\bibitem [{\citenamefont {Popinet}(2003)}]{Popinet2003Gerris}%
  \BibitemOpen
  \bibfield  {author} {\bibinfo {author} {\bibfnamefont {S.}~\bibnamefont
  {Popinet}},\ }\bibfield  {title} {\enquote {\bibinfo {title} {Gerris: a
  tree-based adaptive solver for the incompressible {Euler} equations in
  complex geometries},}\ }\href {\doibase 10.1016/S0021-9991(03)00298-5}
  {\bibfield  {journal} {\bibinfo  {journal} {J. Comput. Phys.}\ }\textbf
  {\bibinfo {volume} {190}},\ \bibinfo {pages} {572--600} (\bibinfo {year}
  {2003})}\BibitemShut {NoStop}%
\bibitem [{\citenamefont {Popinet}(2009)}]{Popinet2009ParasiticCurrents}%
  \BibitemOpen
  \bibfield  {author} {\bibinfo {author} {\bibfnamefont {S.}~\bibnamefont
  {Popinet}},\ }\bibfield  {title} {\enquote {\bibinfo {title} {An accurate
  adaptive solver for surface-tension-driven interfacial flows},}\ }\href
  {\doibase doi.org/10.1016/j.jcp.2009.04.042} {\bibfield  {journal} {\bibinfo
  {journal} {J. Comput. Phys.}\ }\textbf {\bibinfo {volume} {228}},\ \bibinfo
  {pages} {5838--5866} (\bibinfo {year} {2009})}\BibitemShut {NoStop}%
\bibitem [{\citenamefont {Nicholls}\ and\ \citenamefont
  {Ranger}(1969)}]{Nicholls1969AerodynamicShattering}%
  \BibitemOpen
  \bibfield  {author} {\bibinfo {author} {\bibfnamefont {J.~A.}\ \bibnamefont
  {Nicholls}}\ and\ \bibinfo {author} {\bibfnamefont {A.~A.}\ \bibnamefont
  {Ranger}},\ }\bibfield  {title} {\enquote {\bibinfo {title} {Aerodynamic
  shattering of liquid drops},}\ }\href {\doibase 10.2514/3.5087} {\bibfield
  {journal} {\bibinfo  {journal} {AIAA J.}\ }\textbf {\bibinfo {volume} {7}},\
  \bibinfo {pages} {285--290} (\bibinfo {year} {1969})}\BibitemShut {NoStop}%
\end{thebibliography}%
\end{document}